\documentclass{aastex631}

\newcommand{\qstard}{$Q^{*}_{D}$}
\newcommand{\qstarrd}{$Q^{*}_{RD}$}
\newcommand{\pkdgrav}{\texttt{pkdgrav}}

\begin{document}

\title{Satellite formation around the largest asteroids}

\author[0000-0002-0906-1761]{Kevin J. Walsh}
\affiliation{Southwest Research Institute\\
1301 Walnut St. Ste 400\\
Boulder, CO 80305, USA}

\author{Ronald-Louis Ballouz}
\affiliation{Johns Hopkins University Applied Physics Laboratory, Laurel, MD, USA}

\author[0000-0002-3544-298X]{Harrison F. Agrusa}
\affiliation{Universit\'{e} C\^{o}te d’Azur, Observatoire de la C\^{o}te d’Azur, CNRS, Laboratoire Lagrange}

\author{Josef Hanu\u{s}}
\affiliation{Charles University, Faculty of Mathematics and Physics, Institute of Astronomy, V Holes\u{o}vik\'{a}ch 2, 18000 Prague 8, Czech Republic}

\author{Martin Jutzi}
\affiliation{Space Research and Planetary Sciences, Physikalisches Institut, University of Bern, Switzerland}

\author{Patrick Michel}
\affiliation{Universit\'{e} C\^{o}te d’Azur, Observatoire de la C\^{o}te d’Azur, CNRS, Laboratoire Lagrange}
\affiliation{The University of Tokyo, Department of Systems Innovation, Japan}



\begin{abstract}
Satellites around large asteroids are preferentially found among those with the most rapid rotation and elongated shape. The taxonomic statistics are similarly skewed; in total, 13 asteroids larger than 100 km are known to have satellites, but none have been discovered among S-type asteroids. Previous modeling suggests that satellites could be generated by impacts, but spin and shape have never been tracked in models to relate collisional circumstances with those two observed properties concerning the primary. Here we show, by combining simulations of impacts into porous low-density asteroids, their subsequent disruption, reaccumulation and long-term satellite stability, a direct pathway for the formation of satellites. The immediate distortion and elongation of a rotating target body provides a launching point for some debris distinct from simple ballistic ejecta trajectories. The debris that are found to originate from the distorted long-axis is sourced primarily from 10-20 km below the surface and can be placed directly onto eccentric orbits with sufficiently large pericenter distances that avoid rapid re-impact. The specific energy and resultant total mass loss in satellite-forming collisions are not constraining, which explains the observed lack of correlation between asteroids with satellites and those that are part of large asteroid families.

\end{abstract}


\section{Introduction} \label{sec:intro}
The known population of asteroids with satellites is constantly growing with new discoveries. They have been found widely across the various small body populations in the solar system, are being better characterized with new technologies \citep{2021A&A...654A..56V} and have been the subject of {\it in situ} study from recent space missions \citep{2023Natur.616..443D}. There are current space missions that will encounter more in the near future \citep{Brown2021,2023SSRv..219...59N}, while current and future observatories are likely to dramatically increase the list of candidate systems \citep{2024A&A...688A..50L}.

Asteroid satellites provide important information regarding the collisional history of different populations, by way of their existence and properties. Asteroid mass can be measured from satellite’s orbits and when combined with sizes and shapes, which can be constrained with ground-based telescopes, densities can be estimated. Changes in orbital properties of satellites over time can also provide insight into the internal properties of the primary (e.g.  \citealt{2022A&A...657A..76B}). Finally, any differing binary properties between asteroid taxonomic classes may provide insight into the satellite formation process (e.g., \citealt{2023A&A...672A..48M}), the internal properties of different types of large asteroids, or both.

Detection and characterization of asteroid satellites has been done in numerous ways: in situ detection with spacecraft \citep{1995Natur.374..783C,2023Natur.616..443D}, via occultation/eclipses in lightcurve amplitude data \citep{2000Icar..146..556M,1997Icar..127..431P}, direct imaging from ground- or space-based platforms \citep{1999Natur.401..565M} or via radar \citep{2002Sci...296.1445M}. The properties of asteroid satellites have a strong dependence on the primary’s size, where among asteroids with diameter larger than 100 km, satellites are relatively small (\textless 0.1 times the diameter of the primary) and slightly more distant than those found among asteroids smaller than 10 km (see \citealt{2015aste.book..355M,2015aste.book..375W}). There are few known satellites in between, with primaries between 10--100km, where discoveries may be frustrated by observing biases (see \citealt{2023SSRv..219...59N}), although astrometric studies have identified new candidates in this size range yet to be confirmed with other techniques \citep{2024A&A...688A..50L}.

The known satellites around the large asteroids (hereafter ‘large asteroids’ are those with diameter larger than $\sim$100 km) are all on orbits between 3--14 primary radii (Fig. 1), which occupy between ~0.5--4\% of the primary body’s Hill sphere. Where known, their orbital eccentricities are low (below $\sim$0.1). Five of the 13 known systems have multiple satellites with (130) Elektra having three \citep{2022A&A...658L...4B}.  The satellites are relatively small compared to their primary where all of the satellites have radii smaller than $\sim$0.2 primary radii and most are below 0.1 primary radii. Thus, it is expected that more than 99\% of a system’s mass is typically contained in the primary of a multi-asteroid system. 

\begin{figure}[ht!]
  
  \includegraphics[width=0.42\textwidth]{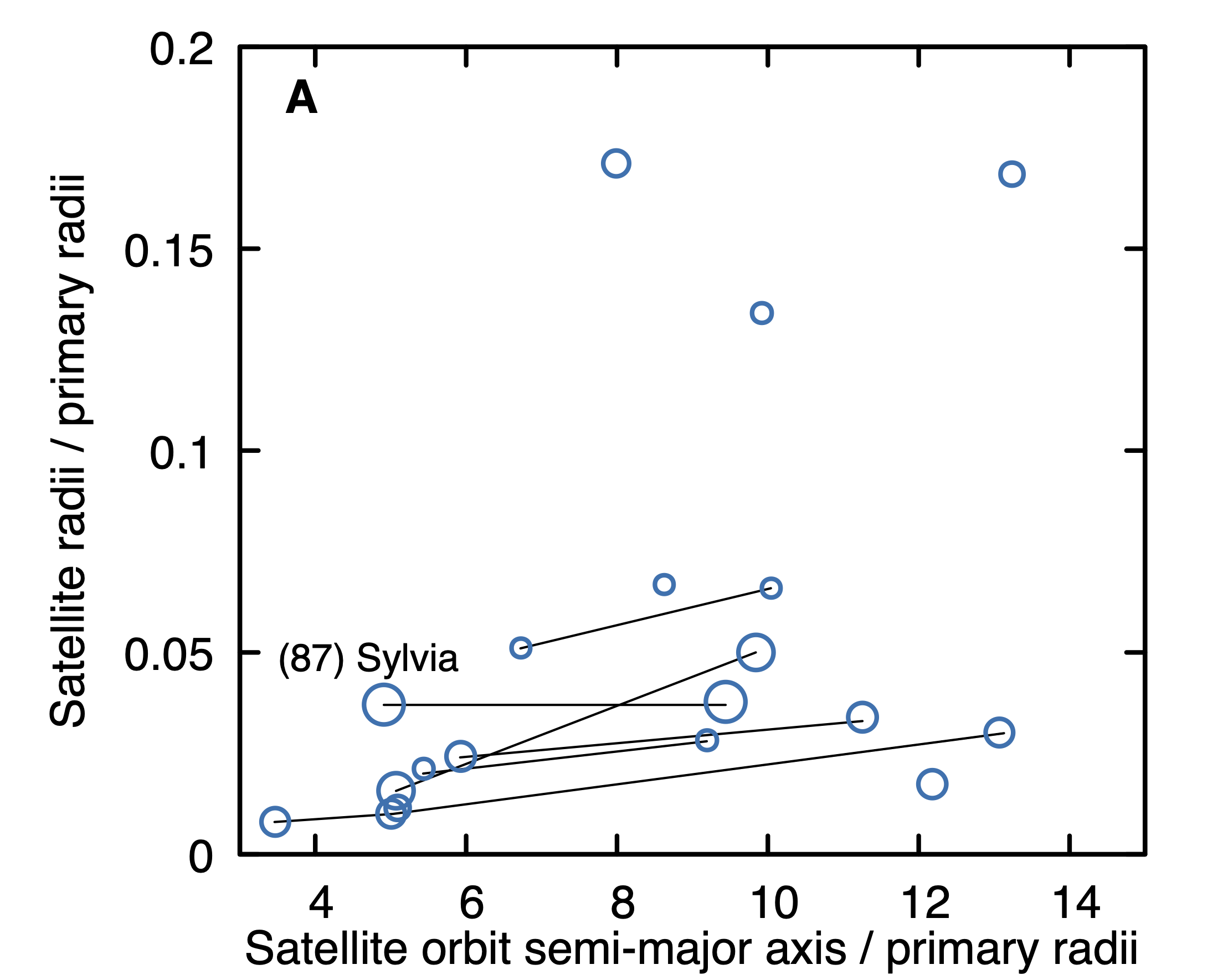}
\includegraphics[width=0.42\textwidth]{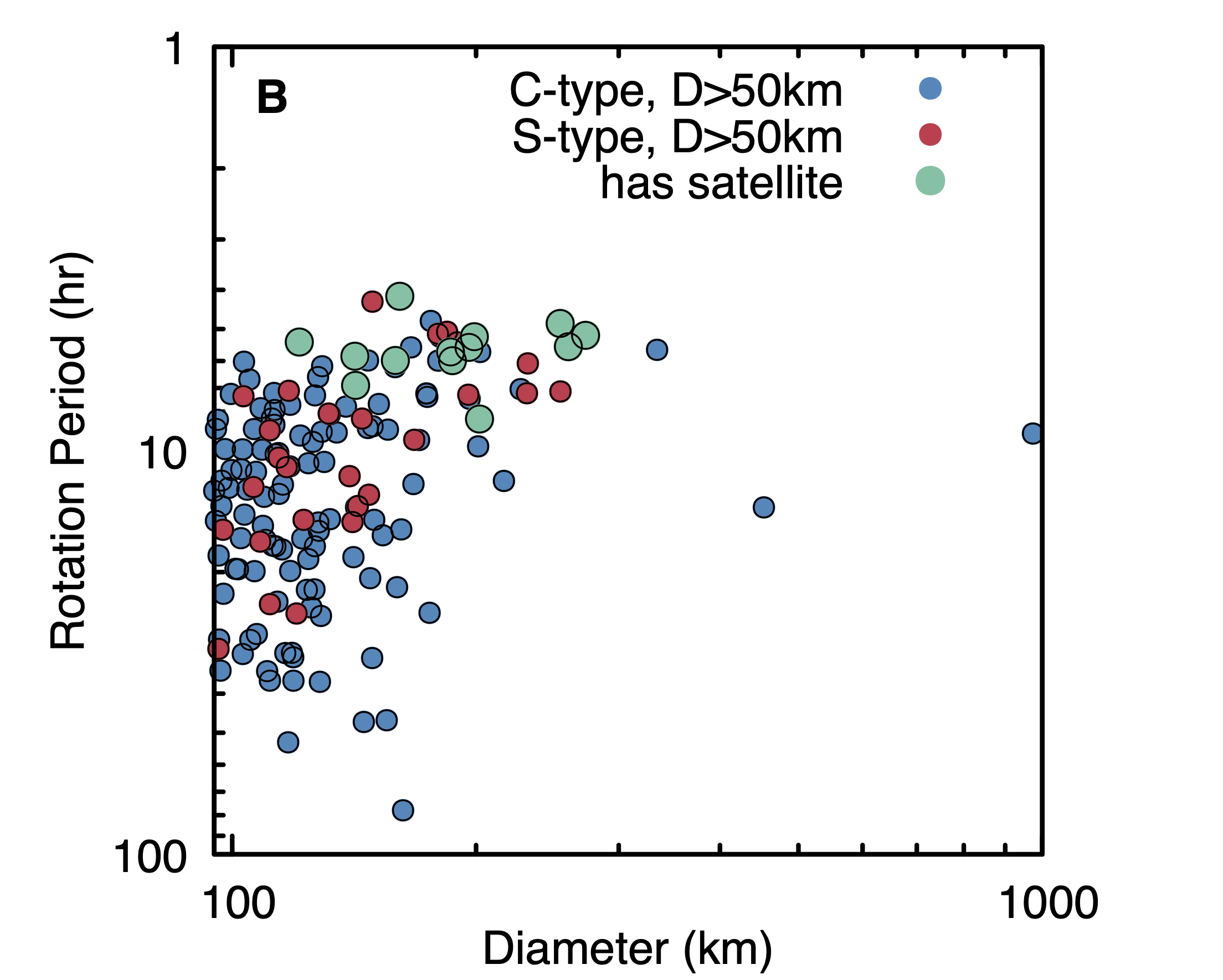}
\includegraphics[width=0.42\textwidth]{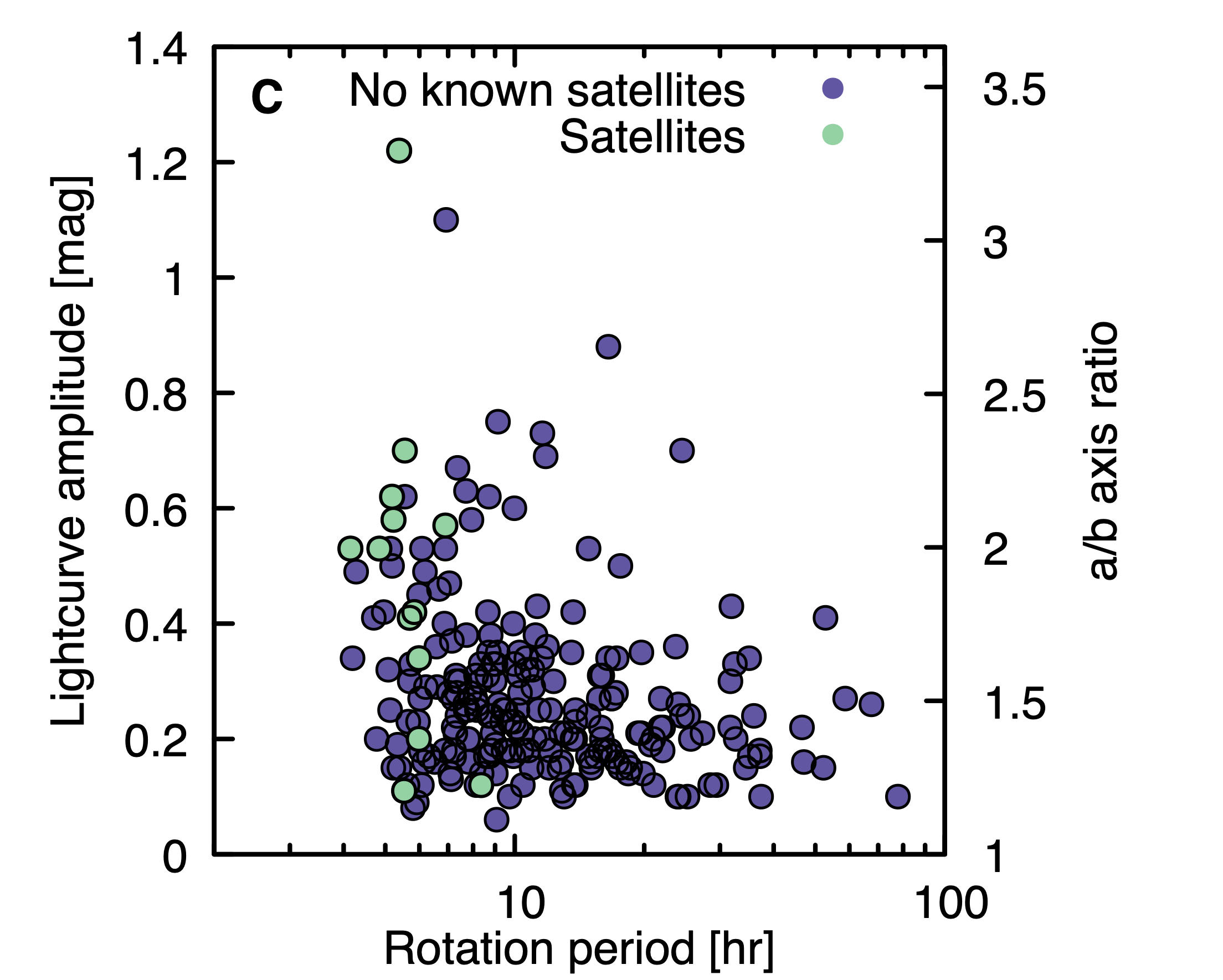}

\caption{The population of observed large asteroids with satellites. (A) The size ratio of satellites to primary body as a function of their orbit semi-major axis scaled by the primary radii for the observed population of satellites among asteroids with diameter greater than 100 km. Multiple systems are connected with lines and the size of each point gives a relative scale of the size of the primary, where $D=286$ km (87) Sylvia is indicated for scale. (B) Rotation periods of asteroids larger than 50 km as a function of diameter derived from lightcurve observations, where S-complex and C-complex objects are in red and blue respectively, and those with observed satellites are aqua. (C) Asteroid lightcurve amplitudes in magnitude plotted as a function of the asteroid’s rotation period in hours where all asteroids are in purple and those with satellites in aqua. The lightcurve amplitudes are from the same lightcurve database and can be roughly translated into axis ratios via: $\Delta m=2.5log(a/b)$, where $\Delta m$ is the lightcurve amplitude and $a$ and $b$ are the long and intermediate principal axes. 
\label{fig:1}}
\end{figure}

The primaries of large main belt asteroid systems are also among the most elongated amongst their peers, as evident from both analysis of lightcurves for a huge number of bodies \citep{2021pds..data...10W} and via direct imaging of $\sim$30 of the largest asteroids \citep{2021A&A...654A..56V}. Lightcurve observations have been published for 222 objects larger than $D>100$ km (with a quality of U=3 that ensures unambiguous period measurements) and they can immediately detect a lower limit for $a/b$ axis ratio (long and intermediate respectively), but require favorable observing geometry over numerous apparitions to constrain the short axis $c$ \citep{2021pds..data...10W}. Direct imaging can provide more detailed shape information, including concavities \citep{2021A&A...654A..56V}. 

\newcommand\z{{\citet{2022A&A...662A..71F}}}
\newcommand\zz{{\citet{2020A&A...643A..38Y}}}
\newcommand\zzz{{\citet{2019A&A...623A.132C}}}
\newcommand\zzzz{{\citet{2021A&A...654A..56V}}}
\newcommand\zzzzz{{\citet{2017A&A...599A..36H}}}
\newcommand\w{{\citet{2017A&A...607A.117V}}}
\newcommand\ww{{\citet{2023A&A...677A.189F}}}
\newcommand\www{{\citet{2021A&A...653A..57M}}}

\begin{table}
\begin{center}
\begin{tabular}{l l l l l l l l l}
Primary & Diameter & \multicolumn{2}{c}{Lightcurve data} & $p_\mathrm{V}$ & Tax & Fam & Dimensions  &\\
                & (km) & per (hr) & ampl                &                &     & &     (km)        \\
\hline
(22) Kalliope   & 150   & 4.14820  & 0.53 & 0.17 & M/x    & Y & 207,145,122 & \z \\
(31) Euphrosyne & 268   & 5.529595 & 0.11 & 0.05 & C      & Y & 286,274,247 & \zz \\
(41) Daphne     & 187   & 5.98798  & 0.34 & 0.08 & Ch     & N & 235,183,153 & \zzz \\
(45) Eugenia    & 188   & 5.699152 & 0.41 & 0.04 & C      & N & 252,191,138 & \zzzz \\
(87) Sylvia     & 274   & 5.183641 & 0.62 & 0.05 & P/X    & Y & 363,249,191 & \zzzz \\
(93) Minerva    & 154   & 5.981768 & 0.20 & 0.07 & C      & N & & \zzzzz \\
(107) Camilla   & 254   & 4.843928 & 0.53 & 0.06 & P/X    & N & & \zzzzz \\
(121) Hermione  & 187   & 5.550877 & 0.70 & 0.08 & Ch     & N & & \w \\
(130) Elektra   & 201   & 5.224663 & 0.58 & 0.09 & Ch     & N & 267,202,151 & \ww \\
(216) Kleopatra & 118   & 5.385280 & 1.22 & 0.15 & M      & N & 267,61,48 & \www \\
(283) Emma      & 142   & 6.89523  & 0.57 & 0.03 & P      & N & & \w  \\
(702) Alauda    & 202   & 8.354    & 0.12 & 0.06 & C      & Y & & \\
(762) Pulcova   & 142   & 5.839    & 0.42 & 0.04 & F/C    & N & & \\  
\end{tabular}
\caption{Data on asteroids with diameters larger than 100 km that have confirmed satellites. The geometric visible albedo values $p_\mathrm{V}$ and taxonomies are taken from the MP3C database. The sizes and rotation periods are based on published shape models derived from disk-resolved images (references in the Notes column). The shape model is unavailable for Alauda and Pulcova, so these parameters are taken from the MP3C database. The "Fam" column indicates whether the asteroid is associated with a collisional family ("Y" for Yes, "N" for No) in which it is assumed to be the largest remnant of the family forming event. Not listed is (90) Antiope, a system of two nearly 90 km components on a close orbit whose combined brightness often leads to it being listed as a body larger than 100 km.}
\label{table:1}
\end{center}
\end{table}

The primaries of known large main belt asteroids are of the taxonomic types C, M and P, with a notable lack of S-type primaries (Table 1). There is also a lack of satellites amongst the very largest asteroids ($D>300$ km) where Ceres and Vesta were both visited by spacecraft with instrumentation capable of finding even very tiny orbiting debris \citep{2018Icar..316..191M}.  Nor have satellites been found around Hygiea or Pallas, the 3rd and 4th largest asteroids, despite being observed with modern ground-based Adaptive Optics imaging \citep{2020NatAs...4..569M,2021A&A...654A..56V}. Meanwhile, asteroids with diameters between approximately 10 and 100 km have a lower frequency of known satellites and less strict correlation between satellite presence and primary shape and spin (Figure 1). Notably, with decreasing primary diameter, satellites with semi-major axes of $\sim$3-14 primary radii would be increasingly hard to spatially resolve with direct imaging techniques \citep{2008ssbn.book..345N}. Therefore, the scarcity of asteroid satellites around these bodies may simply be due to an observational bias, rather than the product of a physical mechanism that would prevent their formation.

It is generally thought that impacts are responsible for creating satellites around large asteroids \citep{2003Natur.421..608M,2004Icar..170..243D}. Impacts are an inescapable event for large asteroids and furthermore they are known to happen, via in situ observations of large impact basins on asteroids such as Vesta \citep{2012Sci...336..694S}, the existence of many tens of families of asteroids \citep{2015aste.book..297N} and collisional probabilities calculated for the well-measured orbital and size distribution of bodies inhabiting the main asteroid belt \citep{2005Icar..179...63B}. Numerical models of collision and reaccumulation have found satellite formation to be possible \citep{2003Natur.421..608M,2007Icar..186..498D,2012Icar..219...57B,2022A&A...664A..69B}. 

Modeling of impact probabilities, the size distribution of asteroids and the energy required to disrupt and disperse half of a target’s mass finds that roughly 4 asteroids larger than 100km should be disrupted every 1 Gyr \citep{2005Icar..179...63B}. Analysis of the main asteroid belt population finds that roughly 10-20 known asteroid families resulted from the disruption of targets larger than 100km \citep{1999Icar..141...65T,2005Icar..179...63B,2013A&A...551A.117B}. However, the largest remnants of very large families do not have observed satellites. Rather, only half of the observed large asteroids with satellites are the largest members of families, all of which are modest or small in size (see  \citealt{2015aste.book..297N}). The lack of correlation between observed satellites and membership in a very large families is suggestive of sub-catastrophic impacts as being more important than catastrophic collisions in the formation of satellites. 

Numerical simulations of the entire satellite formation process are challenging due to the different time and size scales involved in the different phases of the whole collisional process. Modeling the impact shock propagation through the target requires simulation times that scale with the asteroid size and sound speed, on the order of 10-100’s of seconds for 100 km objects. The subsequent ejection and reaccumulation of the debris back onto the largest remnant and into a family of smaller remnants–some of which may end up as bound satellites–requires simulation times on the order of 10’s of hours to days for 100-km objects. Orbital periods of temporary satellites can be many days and their orbits may evolve over many years. Previous large simulation efforts have also attempted to cover a wide range of possible collision scenarios, varying impact speeds, angles, impactor size \citep{2007Icar..186..498D} and parent body’s internal properties (e.g., \citealt{2003Natur.421..608M,2010Icar..207...54J}. The size of the parameter space and the limitation of the computational tools required simplifications such as neglecting the spin and shape of the remnants by performing perfect mergers for colliding particles. These simulations found satellite formation to be more prevalent for disruptions where more than half of the target mass was lost, which occurs when the specific impact energy exceeds the value called the catastrophic disruption threshold, \qstard \citep{2007Icar..186..498D}. Highly oblique impacts with impact angles of $75^\circ$ resulted in no satellites, likely due to impacts that were not very disruptive, with more than 90\% of the mass of the target remaining in the largest remnant in each case tested. 

\section{Results}

We present impact models of asteroid collisions typical of the Main Asteroid Belt with a parameter space designed to explore  direct pathways to generate satellites. Specifically, the parameters are not chosen to cover all parameter space, rather to ensure impact outcomes with a range of final shape and spin state to examine the correlation of satellite production with these post-impact properties. A hydrodynamic model of each impact is performed and the outcome is handed off to N-body gravitational models. The N-body modeling is done with pkdgrav, which is capable of modeling granular mechanical interactions that allows for the evolving shape of the various remnants to be tracked throughout the reaccumulation process. The final shape and spin of the largest remnant is then used in a longer-term dynamical simulation of the initial temporary satellites to determine which, if any, are stable. We simulate head-on and oblique impacts into spherical targets that have a range of pre-impact rotation rates.

\subsection{Simulation framework for hypervelocity impacts into asteroid parent bodies and long-term dynamical evolution of ejecta}

The Smoothed Particle Hydrodynamics (SPH) model is specific to primitive (low-albedo, low density) asteroid classes with compaction as a key physical process \citep{2019Icar..317..215J}. The SPH simulations used an initial porosity of 50\% and bulk densities around 1.3g cm$^{-3}$ which is typical of dark C-complex asteroids \citep{2012P&SS...73...98C}.  The SPH simulations used non-rotating targets and had a resolution of 400,000 particles. 

To retain the post-collision shape and spin of reaccumulated remnants, we utilize a handoff routine between a SPH code \citep{2008Icar..198..242J,2015P&SS..107....3J} that models the impact and shock propagation and an N-body code pkdgrav \citep{2000Icar..143...45R} that handles the subsequent reaccumulation due to self-gravity and the granular mechanics that govern the final shapes and spins of reaccumulated fragments \citep{schwartz2012implementation}. The shape of the largest remnant is captured during the handoff by “wrapping” the SPH particles in a 3D $\alpha$-shape model, which is the outcome of a geometric algorithm that constructs concave shapes around a set of points in Euclidian space \citep{2019MNRAS.485..697B}. This effectively captures the irregular shape of the largest body at this point in time. This is then used to ‘carve’ the same shape from discrete N-body particles \citep{2019MNRAS.485..697B}. The mass and velocity distribution are then transferred to the N-body shape, ensuring linear and angular momentum conservation. The N-body shape is then integrated forward with a soft-sphere discrete element (SSDEM) version of pkdgrav \citep{schwartz2012implementation,2017Icar..294...98Z}. This process is applied only to the largest remnant, which is the primary body for orbiting debris. It has previously been shown that escaping ejecta can be bound to other escaping ejecta (\citealt{2004Icar..170..243D}; “EEBS”), but these are unlikely to include large enough debris to be in the size ranges considered here.  The debris is analyzed to determine whether they are bound to the largest remnant and their instantaneous orbital properties. A more detailed analysis of the evolution of their orbits that considers the final shape and spin of the largest remnant is described below.

To reduce the computational burden, the targets in the SPH models are not rotating, so an additional pre-processing step is completed during the handoff from SPH to N-body models to simulate the effect of a rotating target. This pre-processing step is completed to expand the parameter space of our study with little added computational cost compared to the computational expense of running multiple cases in SPH with varying pre-impact target spin. At the time of handoff, 400 s after the time of impact, an additional change in linear and angular momentum is added to reflect a modeled pre-impact rotation of the primary (see Appendix). This allows an assessment of pre-impact rotation on the outcome of the impact and reaccumulation process. More importantly this produces a wider range of post-impact shape and spin characteristics for the largest remnant allowing a comparison between satellite formation processes and post-impact system properties. 

The SSDEM models have various parameters (static and rolling friction etc.)  that control the outcome of contact and inter-penetration of particles. Here, we use a few simple friction settings previously used in other studies that bracket expectations for asteroid behavior - high friction, low friction and zero friction \citep{2020NatAs...4..569M}. The reaccumulation stage is modeled for 38 hours, during which the largest remnant converges towards its final shape and spin state and the population of initial orbiting debris stabilizes. The final shape is characterized by dynamically equivalent equal-volume ellipsoids (DEEVE; see \citealt{2024PSJ.....5...54A}), where the long semi-axis ($a_\mathrm{pri}$) is typically used to compare with satellites close approach distances and is distinct from the volume-equivalent radius ($R_\mathrm{pri}$).

The SPH simulation resolution of 400,000 means that for a target of diameter 100km, a single simulation particle would have a diameter of 1.4km, which is similar in size to the smallest known satellite in this population (S/2014 (130) 2 orbiting Elektra is 1.6 $\pm$ 0.4 km; \citealt{2022A&A...658L...4B}). At the end of the 38 hr reaccumulation nearly all simulated cases have liberated material that is bound to the largest remnant, and most have tens or hundreds of bound clumps ranging from single particles up to clumps as large as 10$^{-3}$ the mass of the primary. Given the similar size of single simulation particles with known satellites we refer to all orbiting debris as temporary satellites.

The large number of temporary satellites found in nearly all simulations (see Table 2) suggests that additional processes related to temporary satellite survival and orbit evolution must play an important role in removing or combining many or most of them. Therefore, in addition to the impact and reaccumulation models, the medium-term stability (on the order of 1000 days;100’s of orbits) of temporary satellites is also mapped as a function of initial orbits and the shape and spin of the largest remnant. This uses the REBOUND and REBOUNDx simulation packages where the primary body is modeled as a sphere having the same spin rate and $J_2$ and $C_{22}$ gravity terms resulting from the pkdgrav simulations, where these gravity terms represent the degree 2 terms of the primary’s gravity field accounting for its respective oblateness and ellipticity. This modeling provides an estimate of the dynamical lifetimes of the temporary satellites resulting from each simulation.

Thus, the process described above models a nearly catastrophic impact between asteroids, the propagation of the shock through the target and its immediate re-shaping and ejection of debris. Through the first handoff the shape of the largest remnant is maintained and allowed to continue to evolve while its spin may be altered at the handoff and then allowed to continue to evolve over time. The second handoff, from the reaccumulation to long-term stability, freezes in the shape and spin of the largest remnant, which is now the primary body for all bound debris. The sizes of bound debris stop evolving in size, but their orbits continue to evolve. 

\subsection{Sub-catastrophic impact to gravitational reaccumulation of primary and satellites}
The impact scenarios examined are typical of those experienced by large asteroids in the Main Asteroid Belt, where this study used a subset of sub-catastrophic impacts from (38), with impact speeds of 5 km/s over a range of impact angles from 0 to $75^\circ$ by 10-36 km-diameter impactors on to 100 km targets. None of the impacts reached the specific impact energy required to disrupt a target, \qstard (the value where half of the target mass is dispersed). Thus, the modeled scenarios are all sub-catastrophic impacts, with approximately 50-99\% of the total mass returning to the largest remnant. During the handoff to the N-body phase, pkdgrav particles were given additional linear and angular momentum to represent targets with pre-impact spin periods of 3, 4, 6, and 10 hours. The parameters examined were not selected to examine all possible impact scenarios, rather to explore a wide range of post-impact shape and spin configurations of the largest remaining remnant and to permit analysis of satellite formation around them.

The N-body simulations show that the evolution of the system depends heavily on the angular momentum of the system, where cases with pre-impact rotation or high angle of impact often produce outcomes similar to that shown in Figure 2. There is an initial asymmetric ejecta cloud and the primary body is immediately distorted into an elongated shape. In this case a 13 km object impacts a 100 km spherical target with a 10 hr pre-impact rotation at a 60$^{\circ}$ angle at a speed of 5km/s. As a result, the primary is distorted (see Figure 2) and a tail of debris is generated that is distinct from the immediate ejecta cloud that trails the rotation of the primary and begins to clump into discrete remnants.  


\begin{figure}[ht!]
\includegraphics[width=0.85\textwidth]{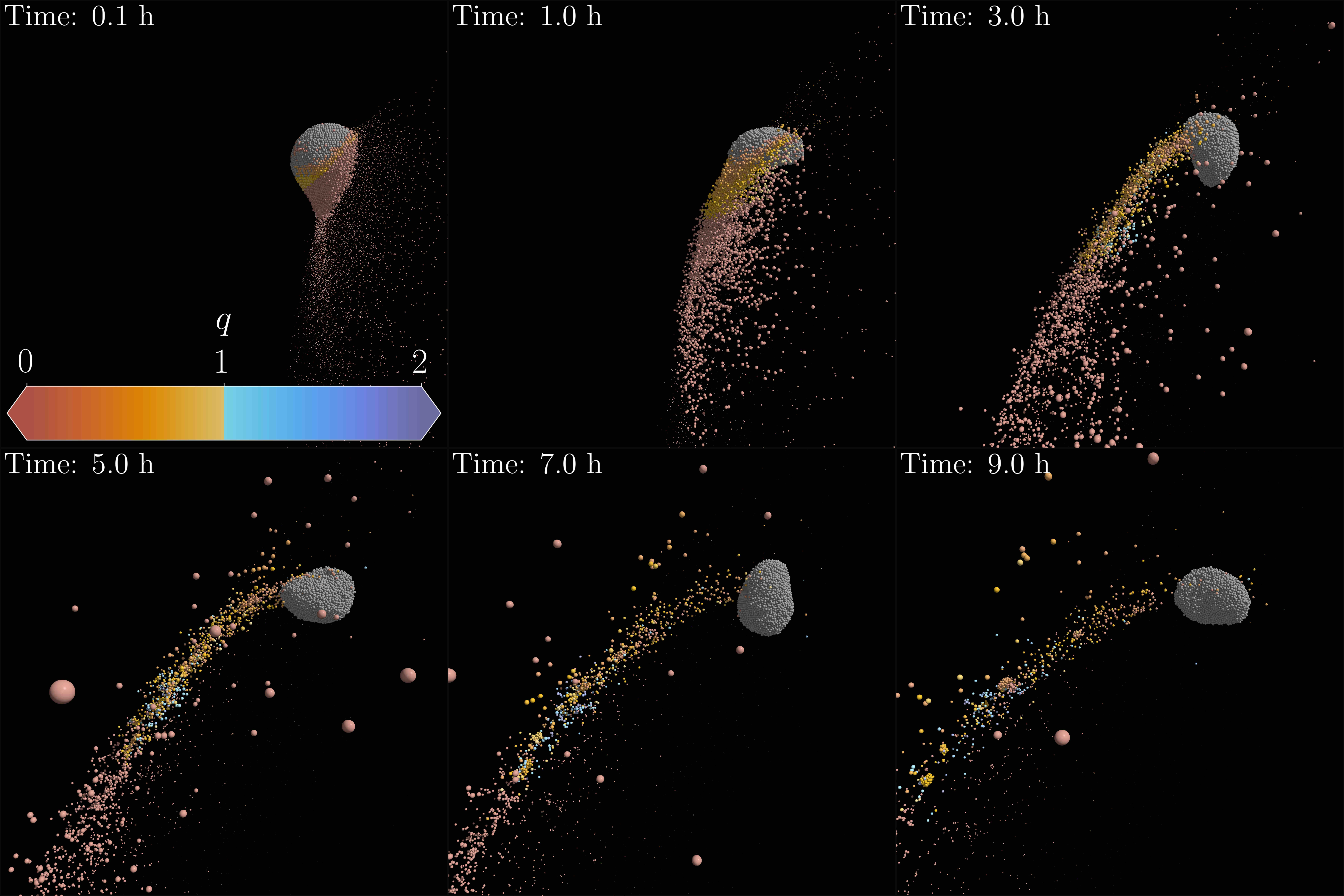}

\caption{Evolution of the target asteroid immediately following impact. The outcome of a 13 km object impacting a 100 km spherical target with a 10 hr pre-impact rotation at a 60$^{\circ}$ angle at a speed of 5 km/s shown at 0.1 h, 1 h, 3 h, 5 h, 7 h and 9 h post-impact. The particles that are part of the main primary mass are colored gray. The rest of the particles are colored by the pericenter of their instantaneous Keplerian orbit. Red colors correspond to trajectories with a pericenter less than the primary’s long axis, while blue colors indicate pericenters above the primary’s long axis. Dark red indicates a pericenter less than zero which corresponds to a hyperbolic (unbound) orbit. See Figures \ref{fig:App_Povray_nospin} and \ref{fig:App_Povray_4hr} in the Appendix for visualizations of the same impact with a non-spinning target and a target initially with a 4 hr rotation period, respectively.
\label{fig:2}}
\end{figure}

Strong correlations are found between the total mass and initial orbital properties of temporary satellites at the end of reaccumulation, specifically related to the shape and spin properties of the primary body.  A search for bound particles and clumps was performed at the end of the reaccumulation phase for each set of parameters, where a hierarchical search for bound companions was performed on the system \citep{2005Icar..176..432L}. Head-on impacts and lower-energy impacts (those that result in minimal re-shaping) result in most ejecta leaving the surface on simple ballistic trajectories that should, without further perturbation, return to their original launch location and thus re-impact. This results in cases where temporary satellites have initial orbit pericenters smaller than the primary’s long axis ($a_\mathrm{pri}$) (Figure 3).

This contrasts sharply with the high angular momentum cases where a significant fraction of the bound ejecta have pericenters larger than the primary’s longest semi-axis, or even twice the longest semi-axis (Figure 3). The debris that reaches the highest pericenter is typically sourced from the tail of debris seen in Figure 2.


\begin{figure}[ht!]
\includegraphics[width=0.42\textwidth]{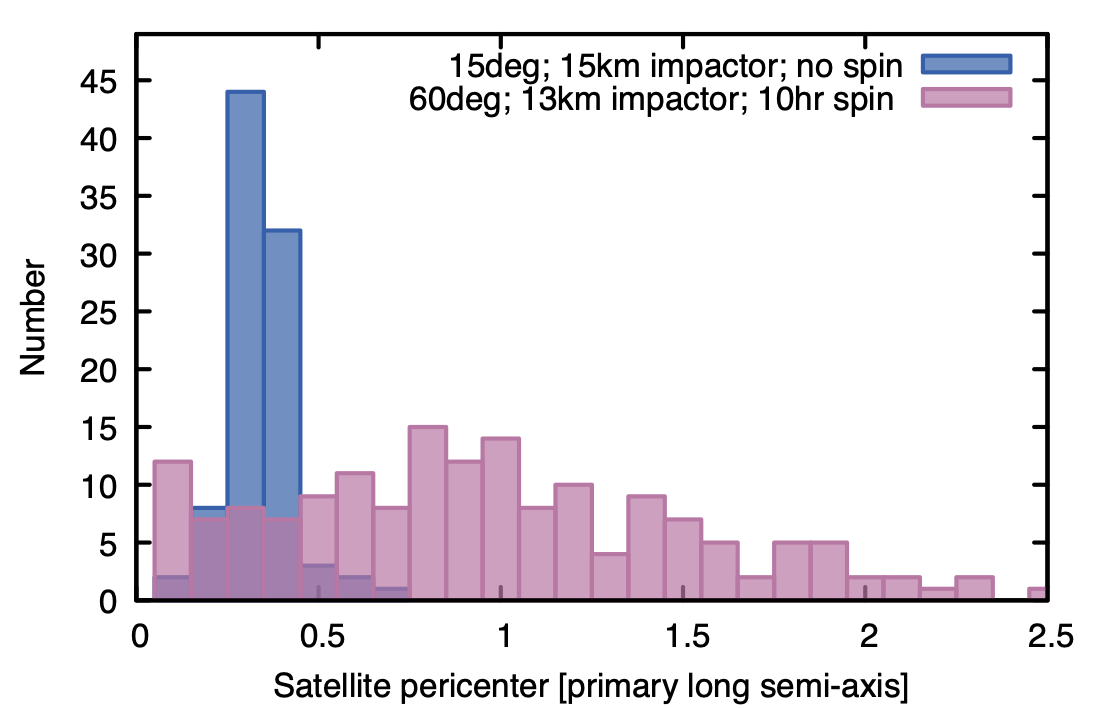}

\caption{Distribution of initial peri-centers for temporary satellites. The distribution of satellite pericenters is shown in blue for a non-spinning target hit at an impact angle of 15$^{\circ}$, where all temporary satellites have initial pericenters below 1 primary radii. This contrasts with a target with moderate spin of 10 hr and a highly oblique impact angle of 60$^{\circ}$, shown in red (the same case as shown in Figure 2), that has significant ejecta on orbits with pericenter above 1 and even above 2 primary long semi-axes lengths.
\label{fig:3}}
\end{figure}

\subsection{Temporary satellite properties}

Observations (Figure 1) also show that there are some large asteroids with elongated shapes, but moderate spin rates (\textgreater 10 hr), that do not have known satellites. Furthermore, the nonspherical gravitational potential around highly elongated primaries will perturb close-in orbits (e.g., \citealt{1999CeMDA..73..339S}). Therefore, it is essential to compare the initial orbits of temporary satellites with stability zones around irregularly-shaped spinning primaries to understand their long-term stability. Likewise, it is essential to compare the total mass of long-term stable satellites with the masses of observed satellites to account for potential accretion processes and minimize dependence on simulation resolution.

At the end of each pkdgrav simulation (38 hr post-impact), the physical and orbital properties of the temporary satellites are tabulated (see Table 2). Specifically, we find that impacts resulting in more elongated primaries, as quantified by the ratio of their major to minor axis ($a/b$), tend to put more mass in orbit (Fig. 4a), and these primaries also tend to have faster post-impact spins (Fig. 4b). Remarkably, we see a sharp transition in the incidence of primaries with satellites at post-impact spin periods of 10 h or less, matching observations. 


\begin{figure}[ht!]
\plotone{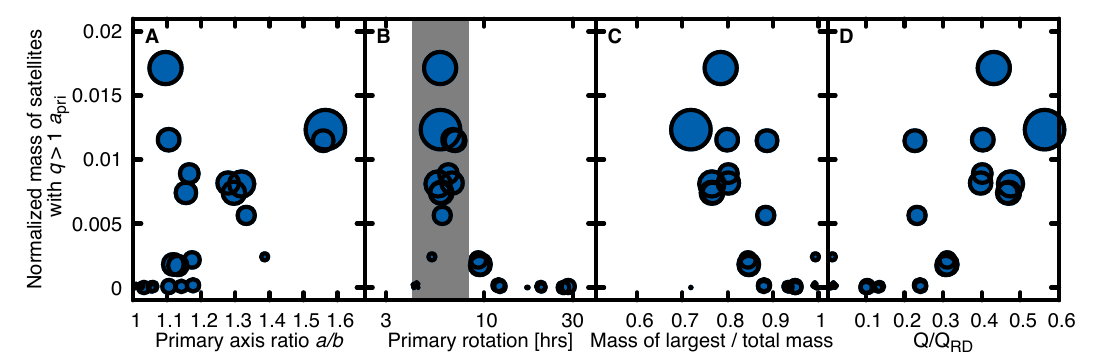}

\caption{Properties of temporary satellites from simulation outcomes. The normalized mass of satellites having initial orbits with peri-centers above 1 primary long semiaxis length as a function of (A) primary post-impact axis ratio $a/b$, (B) primary post-impact rotation period, (C) mass of the largest remnant mass as a fraction of the total system mass (including escaping material) and (D) the ratio of the impact energy to the rotation-dependent catastrophic disruption threshold. The size of each point represents the total number of temporary satellites for each given simulation where the numbers spanned 0--316. The normalized mass is the total mass in satellites as a  fraction of the primary’s mass. The gray zone in the pane B shows the spin rates for all known primaries in the observed sample.
\label{fig:4}}
\end{figure}


\begin{figure}[ht!]
\plottwo{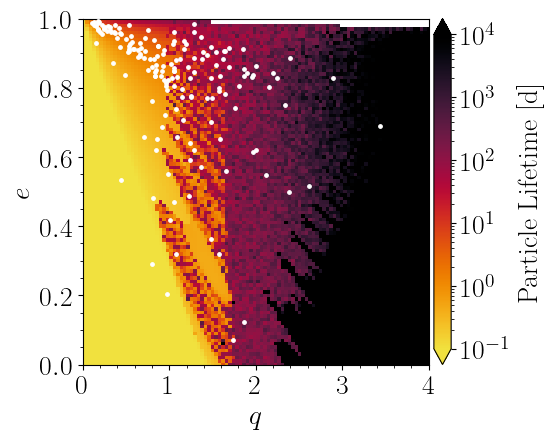}{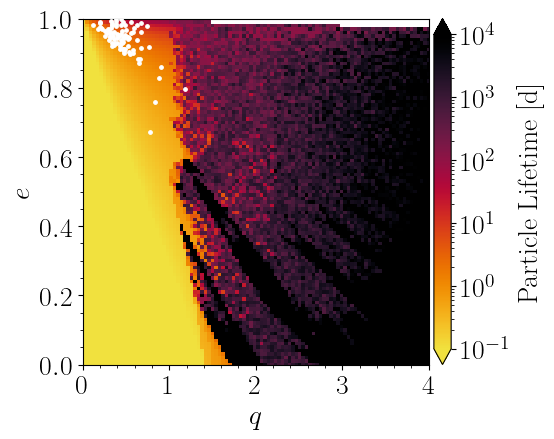}
\caption{Two different impact outcomes with similar post-impact shape and total mass lost, but with different post-impact spin rates, where (A) a case resulting in a mildly elongated final shape ($a,b,c$=57,45,40 km) but rapid 6.7 h rotation period places numerous initial satellites into the medium-term stable regions (satellites plotted as gray dots). The impact scenario had a  13 km impactor hit at an oblique 60$^{\circ}$ impact angle into an initially 10 hr rotating target that resulted in 80\% of the mass remaining in the largest remnant (Case 17 in Table \ref{table:1}).  (B) An outcome with a similarly elongated final shape ($a,b,c$=59,52,34 km) but a final 12.2 h rotation period resulted in nearly all initial orbits having peri-centers below 1 primary radii and almost certain to re-impact. The impactor was 7 km and impacted at a 45$^{\circ}$ angle resulting in a final remnant with 88\% of the total system mass (Case 12 in Table \ref{table:1}). The diagonal structures on these plots are locations of constant semimajor axis and correspond to spin-orbit resonances with the primary.
\label{fig:5}}
\end{figure}

Utilizing the shape and spin of the primary from each simulation, we use REBOUND to compute the lifetime of a test particle on an uninclined orbit as a function of its eccentricity and pericenter distance to constrain the stability of temporary satellites.  Test particles are initialized at apocenter and integrated for 1000 days. The simulation is stopped if the particle impacts the circumscribing sphere of the primary or leaves the primary’s Hill sphere (assuming the primary is located at 2.5 AU). The $J_2$ and $C_{22}$ moments, as well as the primary’s circumscribing sphere are defined based on the body’s moments of inertia at the end of the reaccumulation stage (see Methods). These stability maps find that pericenters beyond $\sim$2-3 primary radii (measured by the primaries’ longest semi-axis length), satellites are stable for numerous orbits (Figure 5).

Not all observed elongated asteroids have satellites. While there are loss mechanisms such as impacts and tides, the stability analyses repeated with elongated shapes but moderate or slow rotation rate finds that the initial stability of ejecta has a direct dependence on the spin rate for asteroids with identical elongated shapes (Figure 6). This effect should act to frustrate satellite formation around large and elongated asteroids that have moderate spin rates. Therefore, we conclude that a combination of both fast rotation and elongation resulting from a large impact is strongly preferred for this direct path to produce stable satellites.


\begin{figure}[ht!]
\plotone{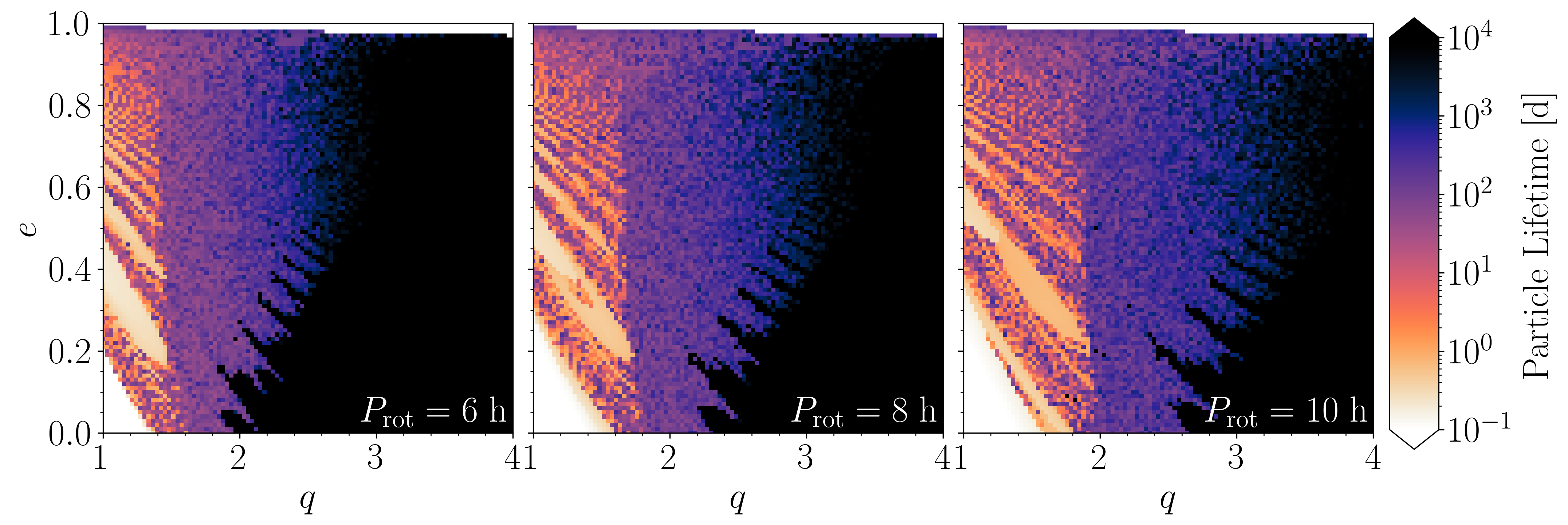}
\caption{Stability maps for different spin periods. Stability maps as a function of rotation period around a primary with a bulk density of 1 g cm$^{-3}$ having the same semiaxes as the primary of Fig. 5a ($a,b,c$ $\sim$ 57,45,40 km) corresponding to a $J_2$ and $C_{22}$ of $\sim$0.05 and $\sim$0.026 when normalized to the primary’s long axis. At short rotation periods, the primary spins much faster than the test particles orbit periods, and the perturbing effect of $C_{22}$ is relatively weak. As the rotation period increases, close-in orbits have shorter lifetimes and the region containing long-term stable orbits is smaller. The colors show the particle lifetime.
\label{fig:6}}
\end{figure}

\subsection{Mass and density effects, S-types, and the very largest asteroids}
There are no known satellites around large S-type asteroids.  There are fewer large S-types in the main asteroid belt ($\sim$30 compared to $\sim$170 non-S-types) providing $\sim$5x fewer targets for large impacts. This still leads to an expectation that a few S-types should have satellites, although it is possible the lack of them is simply an artifact of small number statistics. If S-types required more energy to reach similar levels of disruption owing to their higher density, then an argument could be made about there also being fewer potential impactors. However, two things work against this idea. First, the power law size frequency distribution of asteroids has a slope of approximately -2.5 in this size range, so if twice as much energy is needed to generate an equal disruption it would require an impactor with only 1.25 times larger size. This only decreases the available impactors by a factor of roughly 1.75 relative to C-types. Second, there is no strong evidence for differences in the disruption laws for S-types and C-types where the energy required to disperse the denser S-type target seems to be roughly offset by the extra energy required to crush pore space in lower-density, higher-porosity materials (see \citet{2019Icar..317..215J,2020AJ....160...14B}). 

Another possibility, or contributing factor, is that the initial stability of temporary satellites tends to be lower around a higher density primary body. In general, a primary’s elongated shape (i.e., $C_{22}$) tends to destabilize close-in orbits. But in the limit that the primary’s spin is significantly faster than a satellite’s mean motion, this effect becomes negligible as the perturbing effect averages out over a single orbit period. Therefore, close-in orbits around a fast-rotating, high-density, primary will tend to be less stable due their shorter orbit periods, which are closer to the primary’s spin period. We demonstrate this effect in Fig. 7, where the outcome of a test particle’s orbit is shown as a function of its pericenter distance and eccentricity for a primary with the same elongated shape having different bulk densities and spin periods. The largest stable regions for close-in orbits exist around fast-rotating, low-density primaries, and the size of these regions decrease with increasing density or spin period. The black line in these figures indicate the synchronous orbit, so any orbit below this line would evolve inwards on secular timescales due to tides. In particular, for a high-density primary, the destruction of satellites by tides becomes a more important issue for spin periods longer than shown on these plots. 

\begin{figure}[ht!]
\plotone{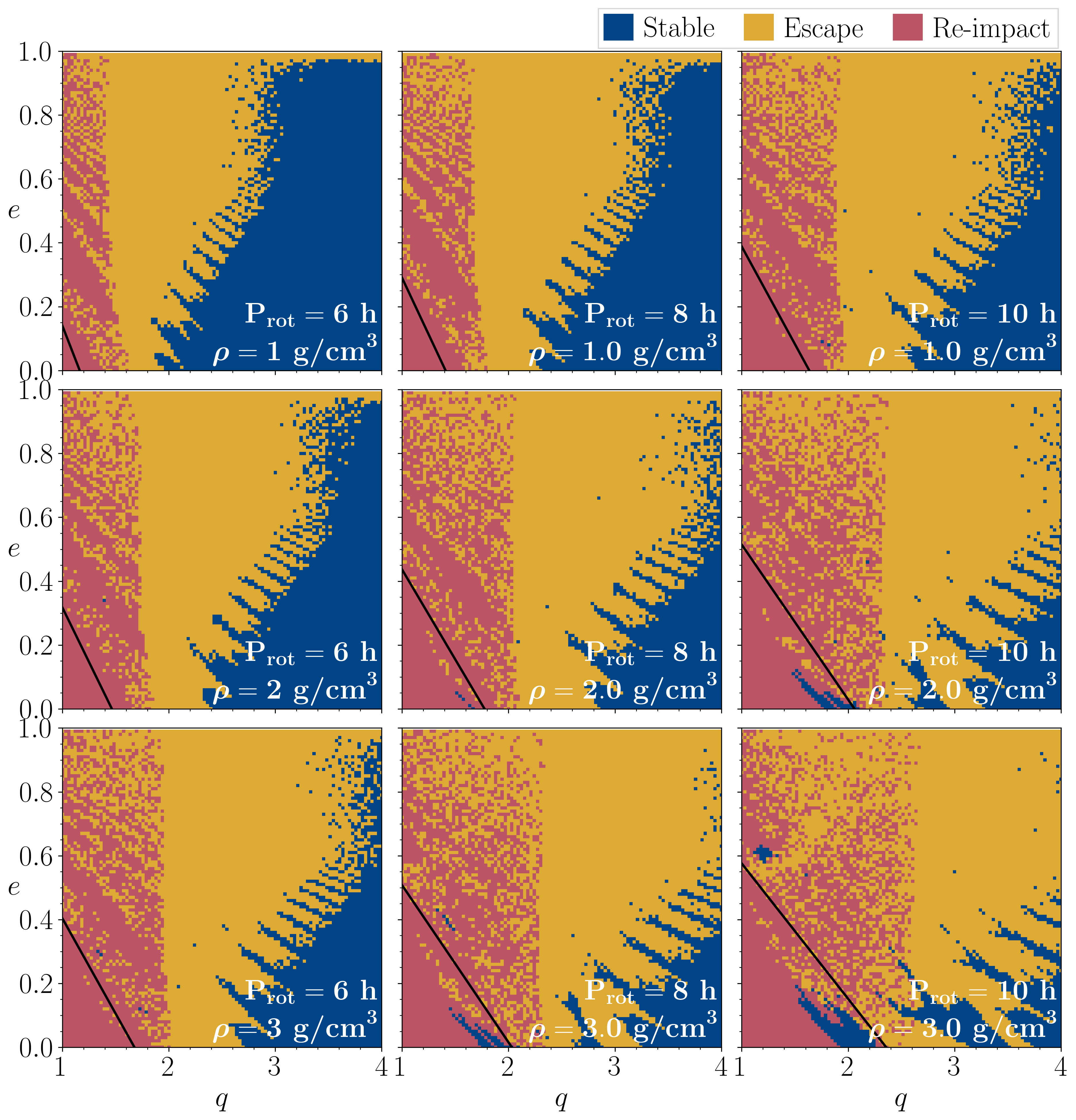}
\caption{The fates of test particles in orbit around an ellipsoidal primary as a function of its bulk density and spin period. The primary has the same semiaxes lengths as those from Figs. 5a and 6 ($a,b,c$ $\sim$ 57,45,40 km) and the top row are the same simulations plotted Fig. 6. The black line indicates the semimajor axis corresponding to the synchronous orbit. Orbits below this line would tidally evolve inwards due to tides. In general, as the primary’s density or spin period is increased, the periods of close-in orbits become closer to the primary’s spin period leading to smaller stable regions owing to the stronger perturbing effect of $C_{22}$.
\label{fig:7}}
\end{figure}

Finally, it is shown here that the correlation between the post-impact shape and spin is what determines likelihood for satellite production and survival. If S-types internal structure or general rheology prohibit or frustrate the fast-spinning outcomes or production of highly elongated shapes then this might explain the lack of observed satellites. The observed distributions of shape and spin for S-types do not indicate an obviously different set of outcomes. A K-S test of the spin frequency distributions and lightcurve amplitude distributions was used to test the null hypothesis that C-type and S-type distributions were drawn from the same distribution. Resultant p-values of 0.06 and 0.28 were found for the spin frequency and amplitude distributions respectively, so we cannot reject the null hypothesis for either and accept that their shape and spin distributions could be similar. 

Notably, some of the known large asteroids with satellites are M-types, with expected, or measured, densities even greater than that typically found for S-types \citep{2012P&SS...73...98C}. These might require even larger impactors to achieve an elongate and fast-spinning outcome. Their properties might speak to a larger class of collisions at a different epoch of Solar System history outside the Main Belt that is discussed later. They may also provide examples of where smaller impacts on targets that gained elongated shapes and rapid rotations through other processes may still provide a pathway to form satellites.

\subsection{Provenance of high-pericenter debris }
Here, the provenance of ejected material that reaches stable orbits is traced back through the modeling to the original SPH target. Outcomes that produce numerous satellites with high pericenter preferentially source material from a few to 10’s km in the subsurface. 

Some cases show discrete depths that preferentially place debris onto high pericenter orbits, suggesting that the specific dynamics of the impact will determine the locations from the target most likely to successfully source the satellites (Figure 8). Specifically, Figure 8 shows four cases with a preferential sourcing of material from depths of 10-20 km. A very disruptive case of a 13 km impactor hitting at 60$^{\circ}$ angle produces this preference for sourcing from depth for all pre-impact spin rates of 3, 4 and 10 hr (Figure 8A,B,C). The effect is strongest at the lowest of those pre-impact spin rates. The same effect is recreated at lower impact angle cases with 4 hr pre-impact spin period (Figure 8D,F). 

The case with the most total temporary satellites, but not the most with high pericenter satellites, sourced the deepest material, with some satellites originating from below 35 km depth (Fig 8C). Meanwhile, for a scenario that produces zero satellites with $q>1.5$ $a_\mathrm{pri}$ all of the temporary satellites were sourced very near the surface (Fig 8E).

\begin{figure}[ht!]
\plotone{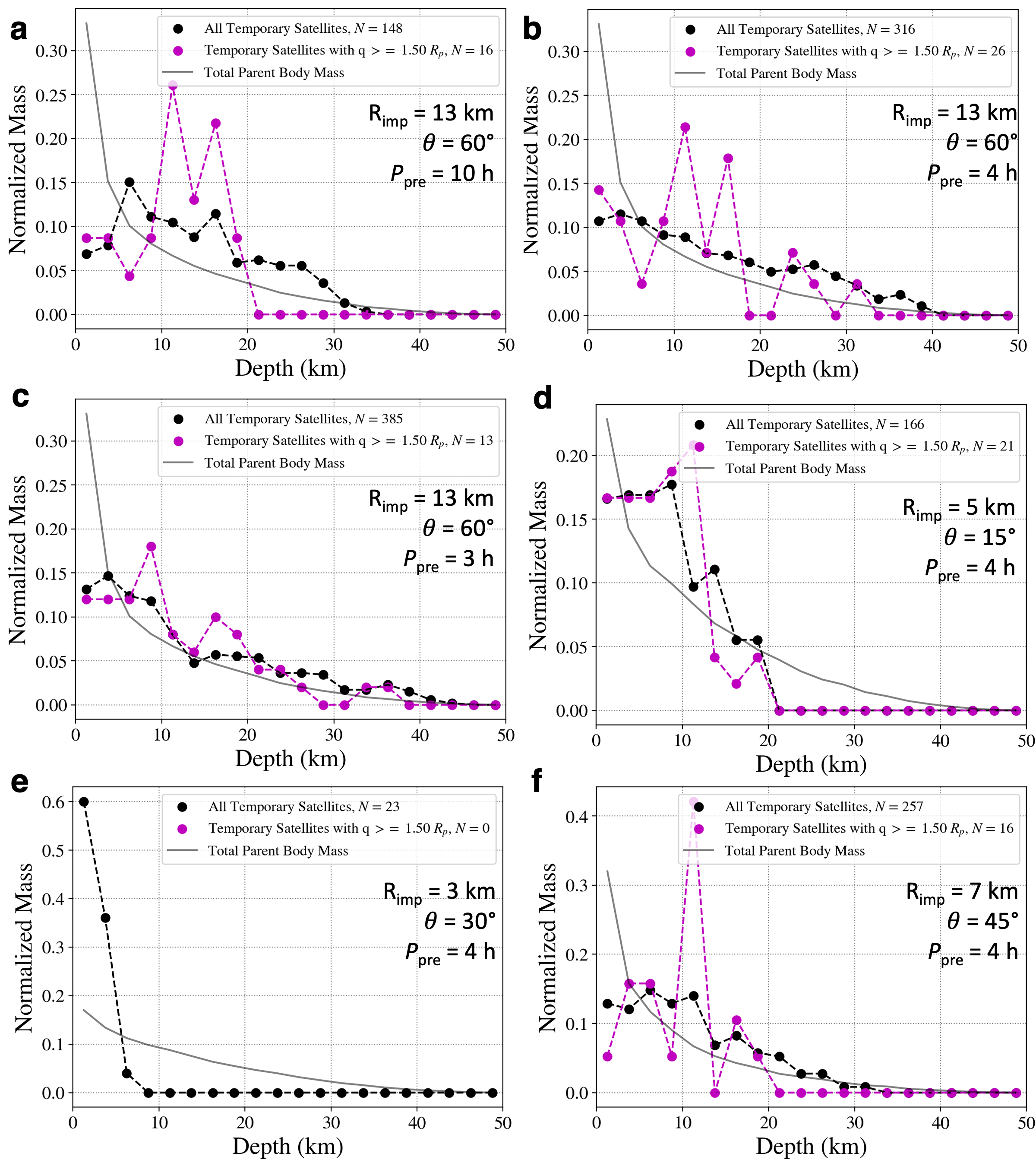}
\caption{Provenance of satellites. The provenance of all temporary satellites (black dots) and temporary satellites with peri-centers greater than 1.5 $R_\mathrm{pri}$  (purple) for six different simulation outcomes. The solid gray line is the distribution of all mass in the parent body and all curves are normalized to show the relative preference for satellite provenance.
\label{fig:8}}
\end{figure}

\section{Discussion}
We presented simulations of asteroid impacts that resulted in a wide range of post-impact shape and spin states. As found in the observed population, there is a correlation between satellite formation and the shape and spin of the largest remaining remnant. The formation of the non-spherical shape, combined with rapid rotation, allows some ejecta to reach orbits that will not immediately re-impact the primary and thus directly reach stable orbits as determined from long-term stability modeling. 

Spherical shapes have the largest regions of stability for orbiting debris, but this work finds that the impacts that produce spherical primaries put minimal or zero ejecta onto initially high pericenter orbits that can remain stable. Meanwhile, the stable zones around elongated asteroids shrink relative to spheres, but the impacts that create such extreme shapes preferentially launch debris onto initially high pericenter orbits. In the simulations presented here, the latter appears to provide an important pathway to satellite formation around large asteroids. This pathway is direct – no additional mechanisms are needed to change or stabilize orbits – and places enough mass in orbit to explain the observed systems.

While the total mass of satellites is similar to that estimated for observed systems, the total number of initially stable satellites is much higher in any given simulation than what is observed, often numbering in the hundreds. Simulation resolution limitations (each particle in these simulations is approximately 1-km in size) probably mean even higher total numbers in reality, or, more likely, the formation of a debris cloud that participates in the formation of a handful stable satellites. Thus, while these simulations are able to reproduce some of the observed characteristics of multi-asteroid systems, additional modeling may be required to fully understand the mechanisms that play a role in the formation and long-term stability of asteroid satellites.

Collisional evolution of the large number of satellites will depend on the orbits and size distribution of debris, where the resolution of the models here are not sufficient to estimate the latter. Collisions are in competition with at least two other effects, where ejecta with close pericenters can be scattered and escape the system due to interactions with the primary \citep{1999CeMDA..73..339S}. The stability simulations presented here aimed to identify ejecta that was stable against scattering, but the 1000 d timescale tested may be shorter than that required for collisional evolution to take over for some cases. If collisions act faster than a satellite can be ejected, then numerous satellites might stabilize and contribute to the population of long-term stable mass. Meanwhile tides will act on very long timescales of 10’s of Myr for orbit circularization and an order of magnitude or longer for orbit expansion \citep{2011Icar..212..661T}. 

If collisional evolution is the dominant evolutionary factor immediately after impact, then ejecta may evolve to a disk without significant loss from dynamical ejection. To explore the end-member case of no dynamical loss, we can make the simple assumption that angular momentum of all of the bound debris is conserved and then a circular equivalent orbit ($a_\mathrm{eq}$; \citealt{2004ARA&A..42..441C}) can be calculated for a resultant relaxed disk. For the systems with significant initial debris with peri-center above 2 primary long semi-axes the typical values for $a_\mathrm{eq}$ are between 4-6 primary radii. If the debris damps to a disk and forms a single, or just a very small number of satellites, then the timescale to tidally evolve their orbits outward could take tens to hundreds of millions of years to evolve outwards towards or beyond 10 primary radii \citep{2011Icar..212..661T}. Notably, nearly all observed satellites are found between 4-14 primary radii. This scenario implies that any individual satellite may include material from numerous initial temporary satellites that went through collisional relaxation to a disk and subsequent accretion.

Catastrophic disruptions that create large families don’t directly correlate with satellite formation. Some, but not all, of the asteroids with satellites are associated with asteroid families and not all of those belonging to families have satellites. Larger asteroid families will be created in cases where less total mass is retained by the target body, which is reported here as the mass of the largest remnant (Figure 4C). The simulation outcomes don’t find a strict correlation between smaller largest remnants (i.e. likely larger asteroid families) and increased mass of stable temporary satellites. Rather there are cases with similar final largest remnants with very different total masses of temporary satellites. 

Instead, we find that the mass of satellites normalized by the primary mass is correlated with the specific impact energy, $Q$, of the collision when it is normalized by a rotation-dependent catastrophic disruption threshold, \qstarrd \citep{2014ApJ...789..158B,2015P&SS..107....3J}. Here, \qstarrd is a simple mapping of the mass loss outcome to the specific impact energy space, through the universal law of catastrophic disruption defined by \citep{2012ApJ...745...79L}. The spread in outcomes for cases with post-impact spin periods $<$ 10 h is due to the fact that those cases represent a wide range in specific impact energies (see Fig 4B and 4D). In the cases of low $Q$/\qstarrd, the impact does not produce temporary satellites; however, the primary’s relatively high pre-impact spin is preserved following the impact. 

With further asteroid discoveries in the Main Belt likely to come from future large surveys it is possible that any large asteroid with a satellite may still be found to have small or dispersed associated families (see \citealt{2022A&A...664A..69B} regarding recent discovery of small family at Kalliope). However, the discovery of very large families is likely close to complete \citep{2013A&A...551A.117B,2021Icar..35714218D}.

There are some primaries that are so large and with such an extreme shape that they challenge the understanding of asteroid family formation and identification, as they have no large detectable family of asteroids associated (this argument goes for asteroids with no known satellites as well). The primary culprit here is Kleopatra, which is the extreme of this group personifying the “dogbone” shape \citep{2016AGUFM.P43B2108M,2021A&A...653A..57M} with a $\sim$5 hr spin rate. Its axes are measured to be 267$\pm$6, 61$\pm$6, 48$\pm$6 km for an astonishing $c/a$ ratio of 0.18 and a clearly double-lobed morphology \citep{2018Icar..311..197S}. Furthermore, the two satellites in orbit have been measured to be tidally evolving in a way that is only possible if they are relatively young ($<$1Gyr) \citep{2021A&A...653A..56B}. Such extreme shapes are not created in the simulations described here and presumably only occur in even more energetic impacts. Such impacts would liberate a huge amount of mass and may require differentiated targets, which are also not considered in our study. Given the very large size of the target asteroid and the large amount of mass that would be lost, such an impact at any point in the solar system history would, presumably, result in a large family of asteroids easily detectable with the numerous family finding algorithms (see \citet{2015aste.book..297N}). Strengthening this case is the young reported age of Kleopatra’s satellites, making it even harder to hide a related enormous asteroid family. 

It has been proposed that some, or all, of the Main Asteroid Belt was implanted from beyond Jupiter or from within the regions where Terrestrial planets formed, inside $\sim$1.5 au \citep{2006Natur.439..821B,2011Natur.475..206W,2017SciA....3E1138R,2024Sci...384..348A}. Therefore, it is possible that an asteroid like Kleopatra suffered its large, re-shaping, impact before it was in the main belt. This general idea is a solution to the “missing mantle” problem where very large and high-density asteroids presumed to have been differentiated asteroids could have lost their mantles in large impacts without littering the entire asteroid belt with their differentiated remnants that are spectroscopically distinct and easy to detect (see \citealt{2015aste.book...13D}). 

If Kleopatra’s satellites are young, this would mean they did not form in the impact that distorted Kleopatra’s shape and left it spinning so rapidly, rather they formed through a different pathway. The large impact, which presumably liberated an enormous amount of material would have happened elsewhere, outside the Main Belt.  Then a later, relatively small, impact would have liberated ejecta that was driven towards stable orbits owing to its existing highly elongated shape and rapid spin. This would rely on a pericenter raising dynamic due to its extraordinary shape and explain the lack of a large associated family and the putative young age of the satellites \citep{2022A&A...657A..76B}. This dynamic is similar to a mass-shedding scenario, postulated for near-Earth Asteroids spun-up by the YORP-effect whose equators are nearly in Keplerian orbits (see  \citealt{2015aste.book..355M,2015aste.book..375W,2021A&A...653A..57M}). Here, we postulate that the extreme shape and spin of Kleopatra provides a pathway for ejecta on simple ballistic trajectories to have their pericenter quickly raised to reach stability. 

The suite of simulations presented here does not test all possible impact conditions or pre-impact shape and spin combinations for targets. In the interest of a manageable parameter space the pre-impact rotation of the target is always prograde with respect to the impactor (see Appendix), the targets are always spherical, and some of the pre-impact rotation rates are among the most extreme observed rotation rates. Rather the parameter space explored here was specifically designed to produce a wide-range of post-impact shape and spin outcomes for the largest remnant of sub-catastrophic impacts. 

Likewise, this work explores only the direct pathway to satellite formation and does not attempt to model the interactions between ejecta, which could, in some cases, lead to collisions and orbit changes even amongst debris initially on $q<1R_\mathrm{pri}$ orbits. Notably, Alauda and Euphrosyne belong to the largest asteroid families amongst the asteroids with known satellites, but they are the least elongated of the observed primaries. A possible solution for these cases is that a large mass of debris was liberated at once in their family-forming impact, giving rise to the chance of ejecta colliding with itself on short timescales and evolving their orbits \citep{2004Icar..170..243D}. The pathways studied here focused on ejecta immediately reaching stable orbits, but a collisional mechanism operating rapidly, could provide the same dynamical tool of raising the pericenter of initially unstable debris, making it long-term stable. Future work focusing on these mechanisms will need to confront the observations that find that not all asteroids that are members of large families have satellites. This suggests that, if the satellites of Alauda and Euphrosyne formed in their family forming event, this process is a relatively low probability outcome.

\begin{acknowledgments}
The authors thank David Minton for a challenging question and important discussion.

This work was supported through numerous funding programs. NASA under grant 80NSSC22K0045 issued through the Solar System Workings Program (KJW, RLB). UCA J.E.D.I. Investments in the Future project managed by the National Research Agency (ANR) with the reference number ANR-15-IDEX-01 (HFA) CNES, The University of Tokyo (PM). The Czech Science Foundation has supported this research through grant 22-17783S (JH). M.J. acknowledges support from SNSF project No. 200021\_207359.

\end{acknowledgments}

\software{\\
\pkdgrav \citep{2000Icar..143...45R}\\
SPH \citep{2019Icar..317..215J}\\
REBOUND \citep{2012A&A...537A.128R,2018MNRAS.473.3351R}\\ 
REBOUNDX \citep{2020MNRAS.491.2885T}
          }



\appendix

\section{Appendix information}
\subsection{Applying pre-impact rotation}
The procedure is as follows: for each particle i, we find the difference in velocity at $t = 0$ (before impact) between the spin rate used in the SPH impact simulation and the desired spin rate for our N-body simulation. As an example, to approximate an impact into a target rotating with period, P, of 6 hr using this procedure, we take the start- and end-state of the SPH simulations of an identical non-rotating target. Then, we map—onto the outcome of the impact into the non- rotating target—the differences in velocities between the rotating and non-rotating targets at $t = 0$. The difference is then added to the particles at the hand off time, $t = t_\mathrm{h}$, for the non-rotating case, providing the conditions for an N-body simulation of a collision onto a rotating target. This procedure can be done for arbitrary values of $P$, as long as individual particles have not experienced many collisions between $t = 0$ and $t = t_\mathrm{h}$. For the simulations presented in this work, this is not the case as the particles are in a ballistic phase before the hand off. The spin of the target is always assumed to be prograde with respect to the angular momentum of the impact.

\begin{figure}[ht!]
\plotone{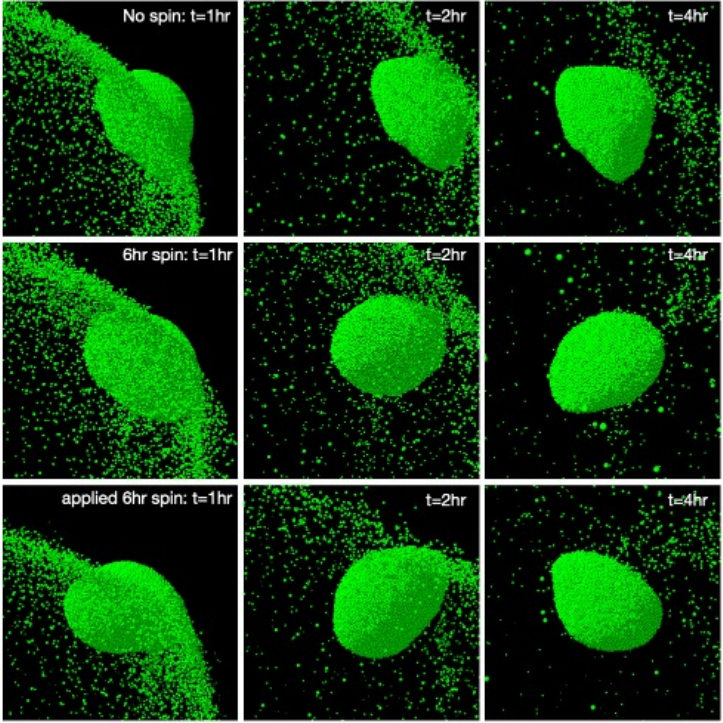}
\caption{The outcome of a 5 km/s impact of a 10 m impactor into a 1 km target with different pre-impact rotation states. All images are following the handoff from SPH to pkdgrav.  A) A non-spinning target results in a moderately prolate object with minimal spin (see Figure 9 for the evolution of the spin rate of the largest remnant). B) A SPH simulation with the target initially spinning at 6h. C) The same impact with the spin adjusted to apply 6 hr pre-impact rotation on the target 100 s after impact.
\label{fig:9}}
\end{figure}

\begin{figure}[ht!]
\plotone{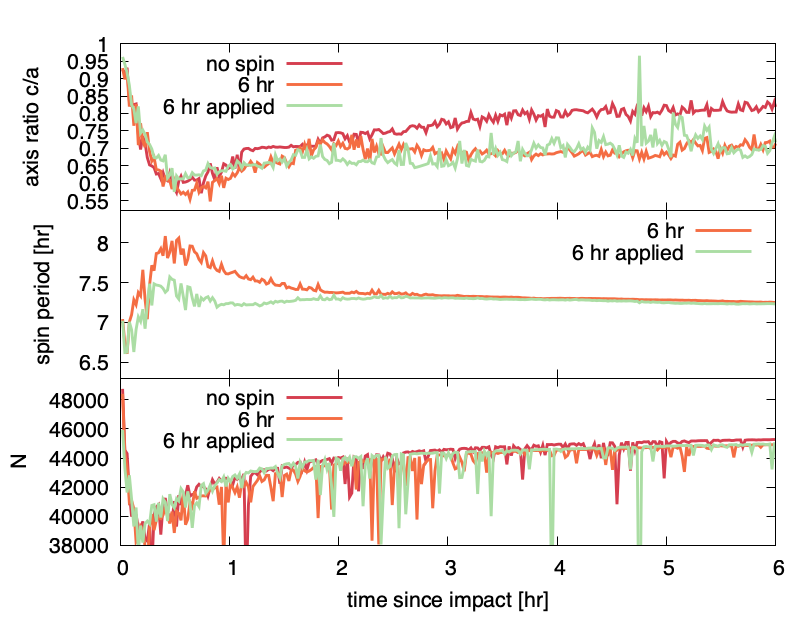}
\caption{The outcome of a 5 km/s impact of a 10 m impactor into a 1 km target with different pre-impact rotation states. A)The axis ratio c/a of the largest remnant as a function of time for all three cases. B) The rotation period of the largest remnant for the 6 hr impact case and the applied 6 hr rotation case (the non-spinning target case converged to 114hr and is not shown). C) The number of N-body particles in the largest remnant over time since impact for all three cases.
\label{fig:10}}
\end{figure}

Here, we provide a comparison of the evolution of the target following the handoff from SPH to pkdgrav for an impact with no pre-impact rotation, one with 6hr pre-impact rotation and one with the prescription applied as described above (Figures \ref{fig:9} and \ref{fig:10}). The impact was at 5 km/s of a 10 m impactor into a 1 km target, and the c/a axis ratio, rotation rate and total accumulated number of particles converge to very similar values for the case of 6 hr pre-impact rotation and 6 hr applied rotation (note in Figure \ref{fig:10} that the rotation rate of the non-spinning target converges to 114 hr and not shown on the figure).

\subsection{Rebound Integration Methods}
To determine the dynamical lifetimes of temporary satellites, we use the REBOUND N-body code (65,66) along with the REBOUNDx package to include the perturbations due to the primary’s $J_2$ and $C_{22}$ moments (67). The $J_2$ and $C_{22}$ moments are defined as (e.g., 68,49,69),

$$J_2 = \frac{C-(A+B)/2}{MR_\mathrm{eq}^2}$$

$$C_{22} = \frac{1}{4}\frac{B-A}{MR_\mathrm{eq}^2}$$

where $M$ is the primary’s mass, $A \leq B \leq C$ are the primary’s moments of inertia, and $R_\mathrm{eq}$ is a normalizing radius which is typically set to the equatorial radius of the primary. We set $R_\mathrm{eq}$ to be the circumscribing sphere of the primary defined by its DEEVE long-axis $a_\mathrm{pri}$ . The primary’s DEEVE semi-axis lengths ($a_\mathrm{pri} \geq b_\mathrm{pri} \geq c_\mathrm{pri}$ ) and moments of inertia are related through the following relations:

$$A=\frac{M}{5}(b_\mathrm{pri}^2 + c_\mathrm{pri}^2)$$
$$B=\frac{M}{5}(a_\mathrm{pri}^2 + c_\mathrm{pri}^2)$$
$$C=\frac{M}{5}(a_\mathrm{pri}^2 + b_\mathrm{pri}^2)$$

The implementation for the primary’s $J_2$ is functionally identical to that already included in REBOUNDx (67). We added the $C_{22}$ term using REBOUNDx, where the additional accelerations felt by a test particle are,

$$\ddot x = -\frac{GM}{r^2}\frac{R_\mathrm{eq}^2}{r^2}\frac{x}{r}C_{22}(15\frac{x^2-y^2}{r^2}-6)$$

$$\ddot y = -\frac{GM}{r^2}\frac{R_\mathrm{eq}^2}{r^2}\frac{y}{r}C_{22}(15\frac{x^2-y^2}{r^2}+6)$$

$$\ddot z = -\frac{GM}{r^2}\frac{R_\mathrm{eq}^2}{r^2}\frac{z}{r}C_{22}(15\frac{x^2-y^2}{r^2})$$

where $x, y, z$, are the particle’s coordinates coordinated in the primary’s body-fixed frame, $r$ is the distance to the primary center, and G is the gravitational constant.

The primary is assumed to be in principal axis rotation with a fixed spin rate equal to the primary’s spin after the pkdgrav simulations (so small rotations about its non principal axes after the pkdgrav simulations are ignored). Particle orbits are integrated using the IAS15 integrator (70). Test particles are initialized at the apocenter of their orbit with varying pericenter distances and integrated forward in time for 1000 days until they either collide with the primary or are ejected from the system. Particles are deemed to have escaped if they pass beyond the systems Hill sphere, assuming the primary is located at 2.5 au. 

\clearpage
\subsection{Table of simulation outcomes}

\begin{table}[hp]
\begin{center}
\begin{tabular}{l l l l l l l l l l l l l}
Simulation ID & $V_\mathrm{imp}$	&	$R_\mathrm{imp}$	&	Angle	&	Friction	&	$P_\mathrm{target}$&	$M_\mathrm{rem}$/	&	$Q/$\qstarrd	&	$P_\mathrm{rem}$	&	$b/a$	&	$c/a$	&	$N_\mathrm{sat}$	&	$N_\mathrm{sat}$	\\

 & [km/s]	&	[km]	&	[deg]	&	[deg]	&	[hr]&	$M_\mathrm{target}$	&		&	[hr]	&		&		&		&	$(q>1a_\mathrm{pri})$	\\
\hline
1 & 5   &   3   &   15  &   18  &   4           &  0.99	&	0.02	&	4.37	&	0.95	&	0.92	&	27	&	4	\\
2 & 5	&	3	&	15	&	18	&	$\infty$	&	0.99 & \textless0.01 & 84.7 & 0.98 & 0.89 & 0 & 0  \\
3 & 5	&	5	&	15	&	18	&	4	&	0.89	&	0.23	&	7.01	&	0.64	&	0.51	&	166	&	85	\\
4 & 5	&	5	&	15	&	18	&	$\infty$	&	0.95	&	0.1	&	26.7	&	0.97	&	0.72	&	92	&	0	\\
5 & 5	&	3	&	30	&	18	&	4	&	0.99	&	0.02	&	4.28	&	0.97	&	0.94	&	23	&	2	\\
6 & 5	&	3	&	30	&	18	&	$\infty$	&	0.99	&	\textless0.01	&	47.3	&	0.99	&	0.92	&	12	&	0	\\
7 & 5	&	5	&	30	&	18	&	4	&	0.88	&	0.23	&	5.98	&	0.75	&	0.52	&	137	&	55	\\
8 & 5	&	5	&	30	&	18	&	$\infty$	&	0.93	&	0.13	&	20.36	&	0.95	&	0.59	&	67	&	0	\\
9 & 5	&	3	&	45	&	18	&	4	&	0.99	&	0.02	&	4.21	&	0.99	&	0.97	&	18	&	3	\\
10 & 5	&	3	&	45	&	18	&	$\infty$	&	0.99	&	0.01	&	44.5	&	0.99	&	0.94	&	13	&	0	\\
11 & 5	&	7	&	45	&	18	&	4	&	0.78	&	0.43	&	5.83	&	0.91	&	0.54	&	257	&	160	\\
12 & 5	&	7	&	45	&	18	&	$\infty$	&	0.88	&	0.24	&	12.2	&	0.88	&	0.57	&	91	&	1	\\
13 & 5	&	3	&	60	&	18	&	4	&	0.72	&	\textless0.01	&	4.22	&	0.99	&	0.98	&	11	&	0	\\
14 & 5	&	3	&	60	&	18	&	$\infty$	&	0.99	&	\textless0.01	&	51.57	&	0.99	&	0.98	&	10	&	0	\\
15 & 5	&	13	&	60	&	18	&	4	&	0.72	&	0.56	&	5.86	&	0.64	&	0.49	&	316	&	120	\\
16 & 5	&	13	&	60	&	18	&	6	&	0.76	&	0.47	&	5.91	&	0.77	&	0.65	&	177	&	74	\\
17 & 5	&	13	&	60	&	18	&	10	&	0.8	&	0.4	&	6.75	&	0.78	&	0.69	&	170	&	81	\\
18 & 5	&	13	&	60	&	18	&	$\infty$	&	0.85	&	0.31	&	9.48	&	0.88	&	0.62	&	154	&	18	\\
19 & 5	&	3	&	75	&	18	&	4	&	0.99	&	\textless0.01	&	4.45	&	0.98	&	0.97	&	4	&	1	\\
20 & 5	&	3	&	75	&	18	&	$\infty$	&	0.99	&	\textless0.01	&	191.4	&	0.99	&	0.98	&	4	&	0	\\
21 & 5	&	18	&	75	&	18	&	4	&	0.99	&	0.01	&	5.27	&	0.72	&	0.58	&	57	&	36	\\
22 & 5	&	18	&	75	&	18	&	$\infty$	&	0.99	&	\textless0.01	&	17.22	&	0.98	&	0.88	&	12	&	0	\\
	&		&		&		&		&		&		&		&		&		&		&		\\
23 & 5	&	13	&	60	&	33	&	6	&	0.76	&	0.47	&	5.68	&	0.76	&	0.67	&	202	&	85	\\
24 & 5	&	13	&	60	&	33	&	10	&	0.8	&	0.4	&	6.45	&	0.86	&	0.65	&	149	&	83	\\
25 & 5	&	13	&	60	&	33	&	$\infty$	&	0.85	&	0.31	&	9.59	&	0.9	&	0.56	&	165	&	19	\\
	&		&		&		&		&		&		&		&		&		&		&		\\
26 & 5	&	5	&	15	&	0	&	$\infty$	&	0.95	&	0.1	&	28.34	&	0.91	&	0.85	&	105	&	1	\\
27 & 5	&	5	&	30	&	0	&	$\infty$	&	0.93	&	0.14	&	20.21	&	0.95	&	0.8	&	75	&	1	\\
28 & 5	&	7	&	45	&	0	&	$\infty$	&	0.88	&	0.24	&	12	&	0.85	&	0.82	&	98	&	2	\\
29 & 5	&	13	&	60	&	0	&	6	&	0.77	&	0.47	&	5.85	&	0.87	&	0.64	&	164	&	76	\\
30 & 5	&	13	&	60	&	0	&	10	&	0.8	&	0.4	&	6.84	&	0.91	&	0.73	&	172	&	94	\\
31 & 5	&	13	&	60	&	0	&	$\infty$	&	0.84	&	0.31	&	9.35	&	0.85	&	0.78	&	123	&	22	\\
32 & 5	&	18	&	75	&	0	&	$\infty$	&	0.99	&	\textless0.01	&	17	&	0.98	&	0.89	&	11	&	0	\\
\end{tabular}
\caption{Simulation outcomes for the model parameters of impact speed ($V_\mathrm{imp}$), impactor radius ($R_\mathrm{imp}$), impact angle, the angle of friction of the \pkdgrav simulation settings and the pre-impact rotation of the target ($P_\mathrm{target}$). The simulation outcomes for each impact are the mass of the largest remnant as a fraction of the target mass ($M_\mathrm{rem}$/$M_\mathrm{target}$), the impact energy scaled by the rotation-dependent catastrophic disruption threshold ($Q$/\qstarrd), the rotation period of the largest remnant, its intermediate over long axis ratio ($b/a$) and its short over long axis ratio ($b/a$), the number of total temporary satellites formed ($N_\mathrm{sat}$) and the number with orbit pericenter larger than the long semiaxes length.}
\end{center}
\end{table}

\clearpage

\begin{figure}[ht!]
\plotone{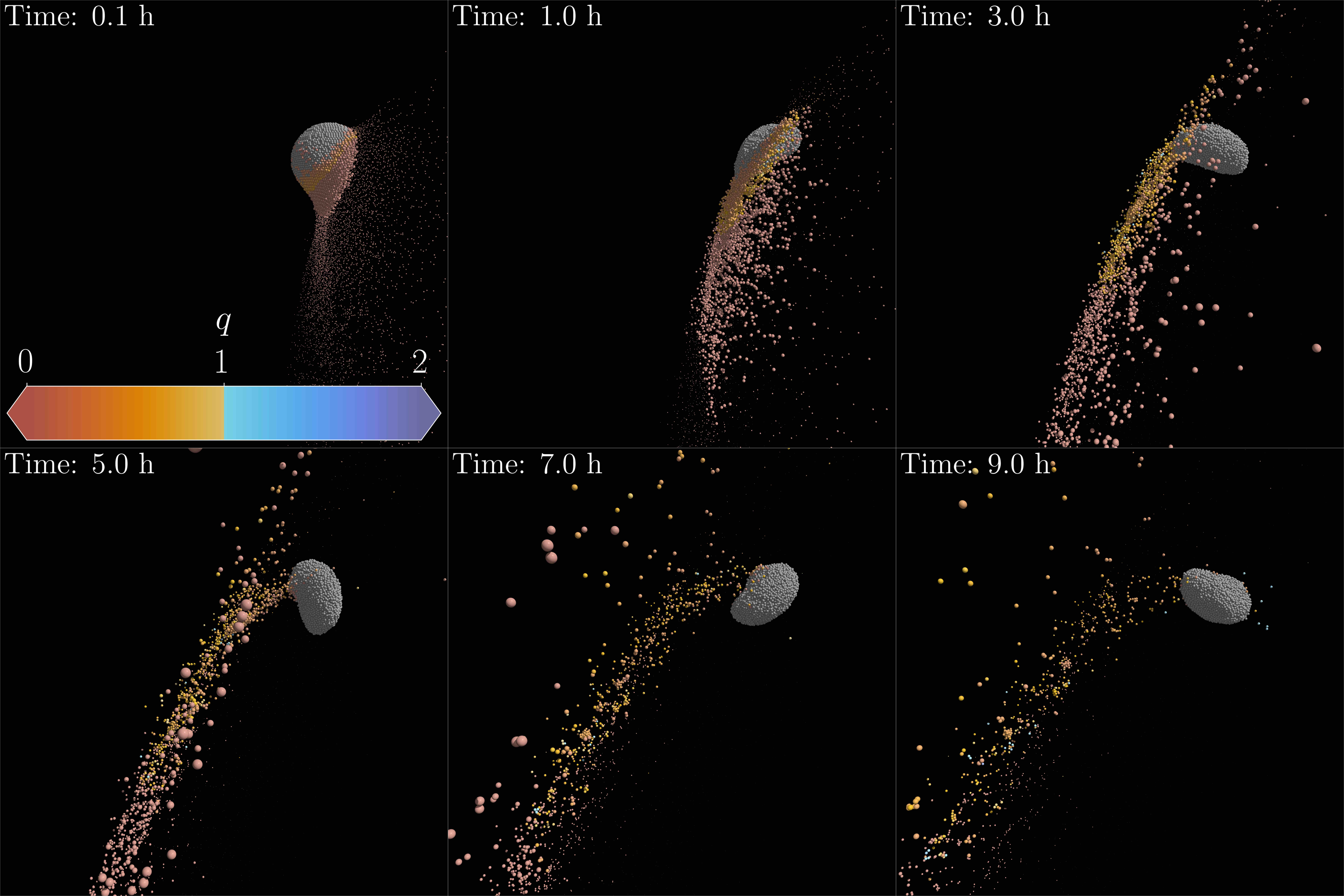}
\caption{Evolution of the target asteroid immediately following impact. The outcome of a 13 km object impacting a 100 km spherical target with no rotation at a 60$^{\circ}$ angle at a speed of 5 km/s shown at 0.1 h, 1 h, 3 h, 5 h, 7 h and 9 h post-impact. The particles that are part of the main primary mass are colored gray. The rest of the particles are colored by the pericenter of their instantaneous Keplerian orbit. Red colors correspond to trajectories with a pericenter less than the primary’s long axis, while blue colors indicate pericenters above the primary’s long axis. Dark red indicates a pericenter less than zero which corresponds to a hyperbolic (unbound) orbit.
\label{fig:App_Povray_nospin}}
\end{figure}

\begin{figure}[ht!]
\plotone{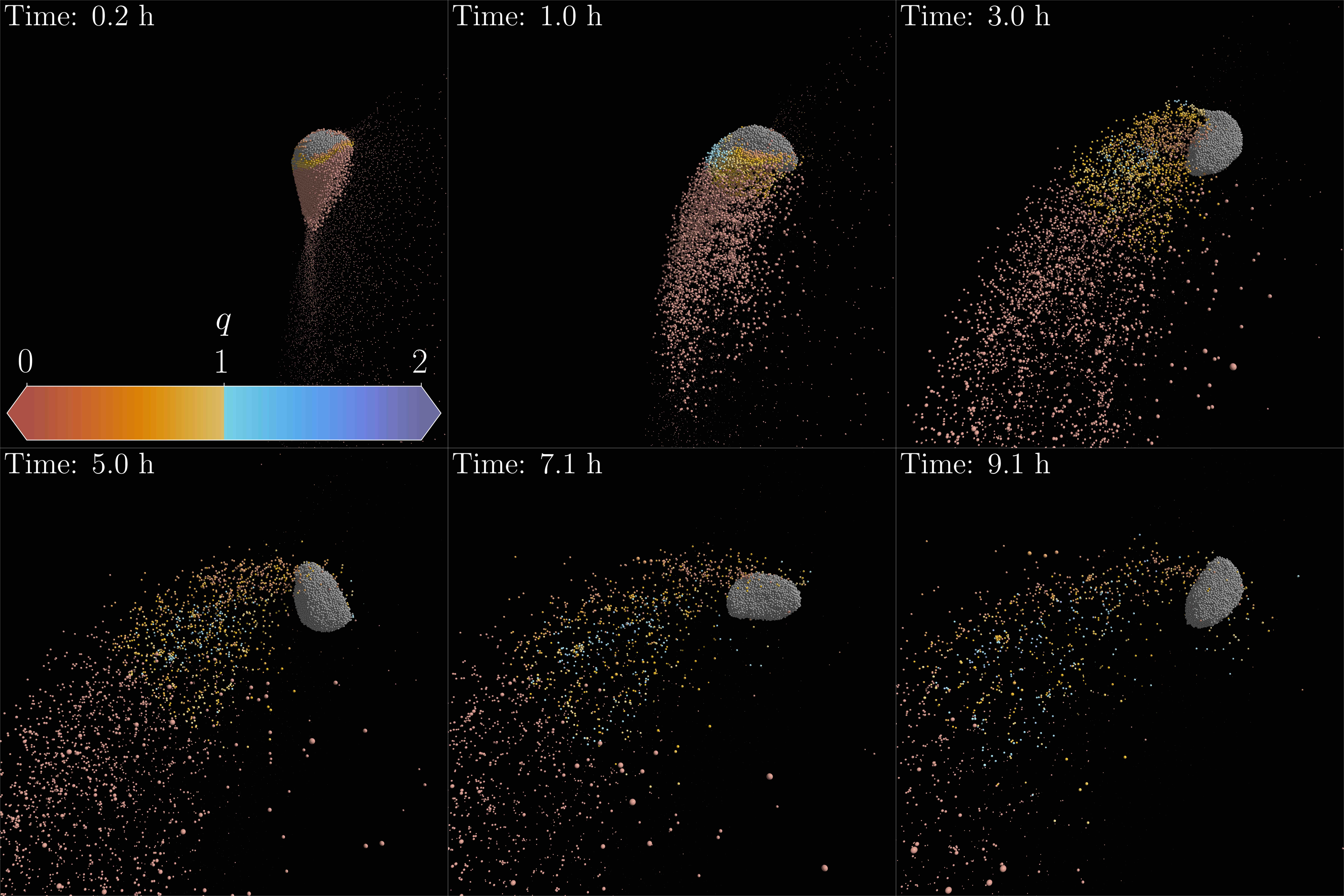}
\caption{Evolution of the target asteroid immediately following impact. The outcome of a 13 km object impacting a 100 km spherical target with a 4 hr pre-impact rotation at a 60$^{\circ}$ angle at a speed of 5 km/s shown at 0.1 h, 1 h, 3 h, 5 h, 7 h and 9 h post-impact. The particles that are part of the main primary mass are colored gray. The rest of the particles are colored by the pericenter of their instantaneous Keplerian orbit. Red colors correspond to trajectories with a pericenter less than the primary’s long axis, while blue colors indicate pericenters above the primary’s long axis. Dark red indicates a pericenter less than zero which corresponds to a hyperbolic (unbound) orbit.
\label{fig:App_Povray_4hr}}
\end{figure}

\clearpage

\subsection{Stability plots for all simulated impacts}

\begin{figure}[hp]
    \gridline{
        \fig{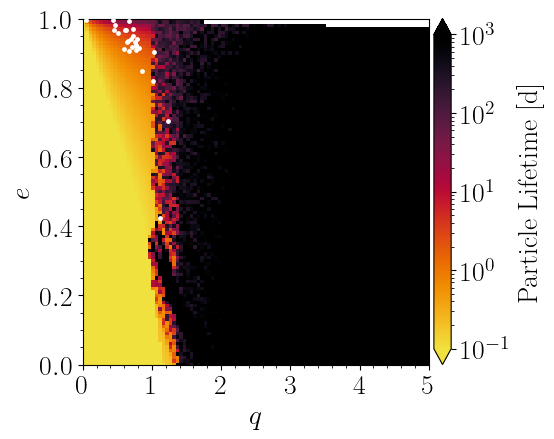}{0.25\textwidth}{Case 1} 
        \fig{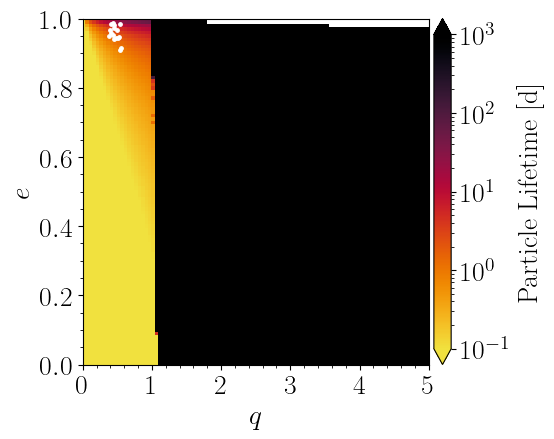}{0.25\textwidth}{Case 2}
        \fig{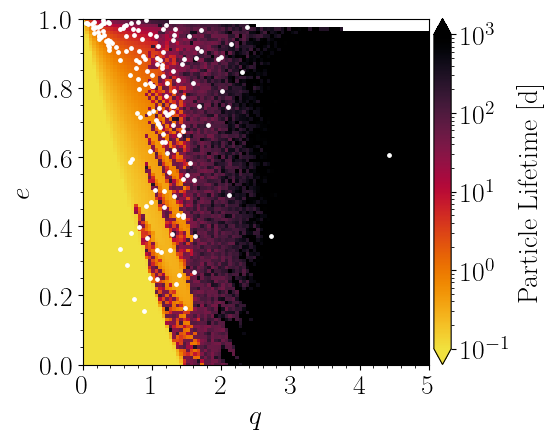}{0.25\textwidth}{Case 3}
        \fig{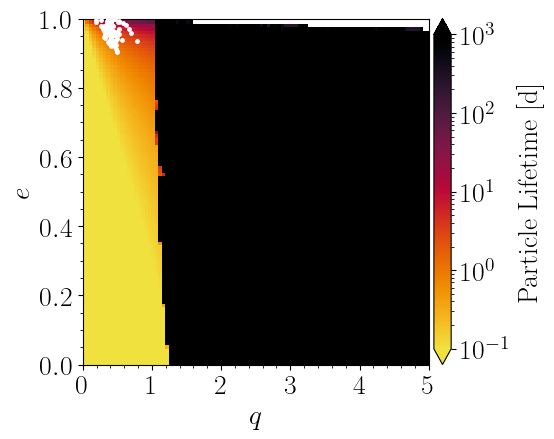}{0.25\textwidth}{Case 4}
    }

    \gridline{
        \fig{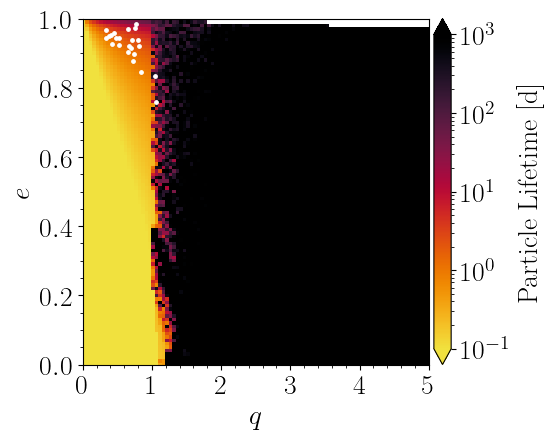}{0.25\textwidth}{Case 5} 
        \fig{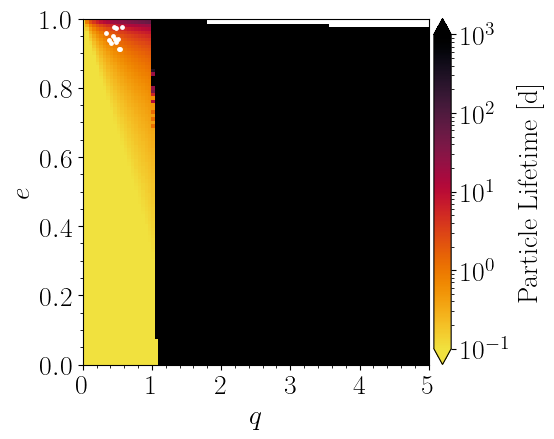}{0.25\textwidth}{Case 6}
        \fig{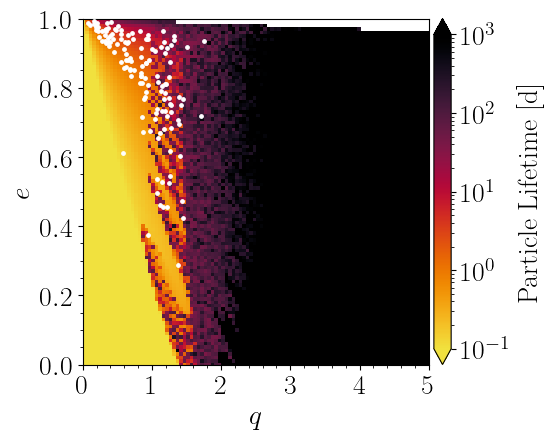}{0.25\textwidth}{Case 7}
        \fig{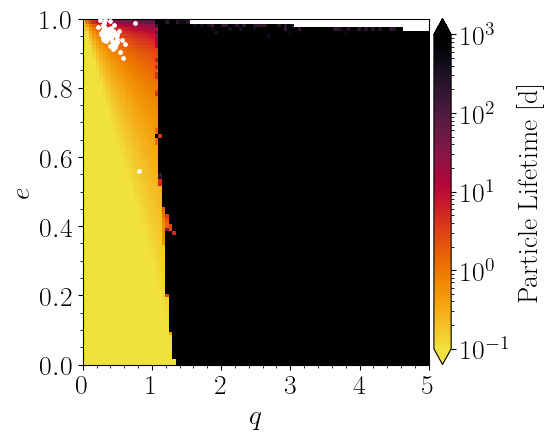}{0.25\textwidth}{Case 8}
    }

    \gridline{
        \fig{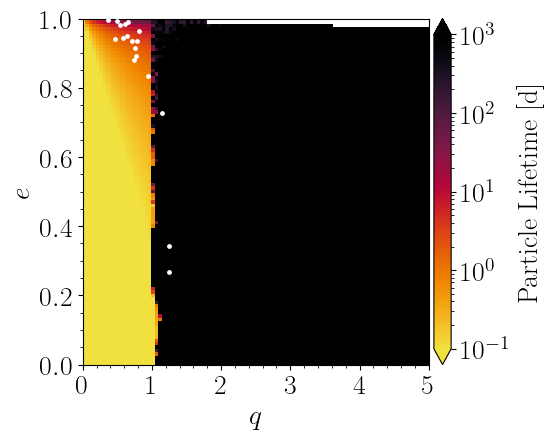}{0.25\textwidth}{Case 9} 
        \fig{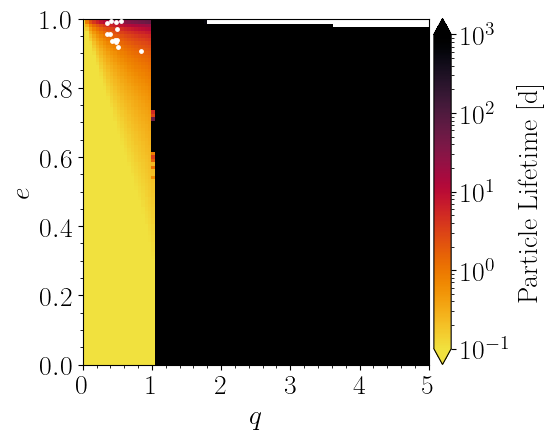}{0.25\textwidth}{Case 10}
        \fig{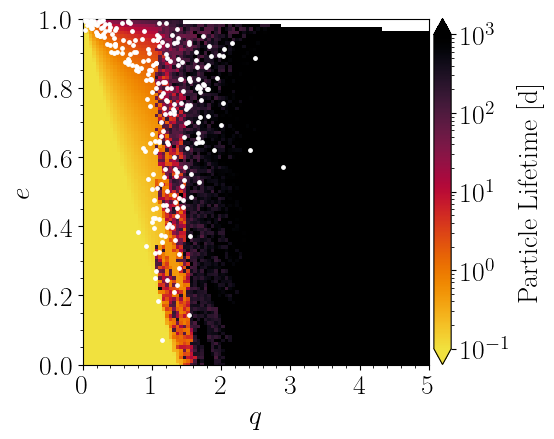}{0.25\textwidth}{Case 11}
        \fig{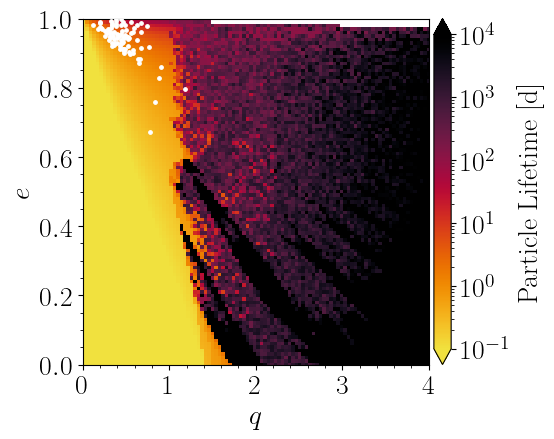}{0.25\textwidth}{Case 12}
    }

    \gridline{
        \fig{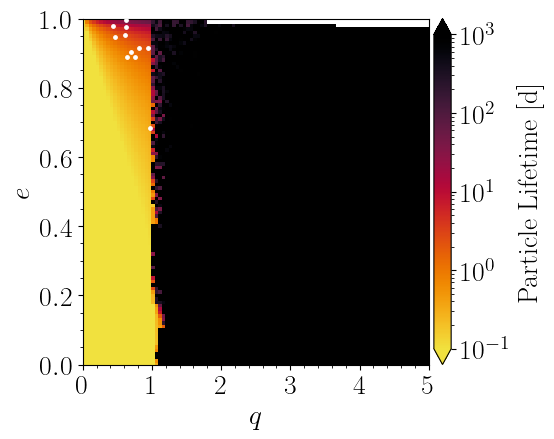}{0.25\textwidth}{Case 13} 
        \fig{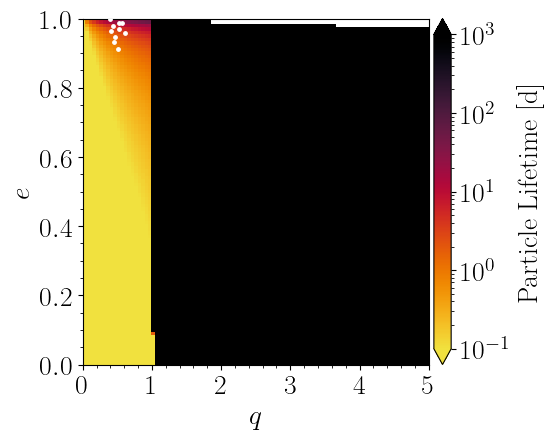}{0.25\textwidth}{Case 14}
        \fig{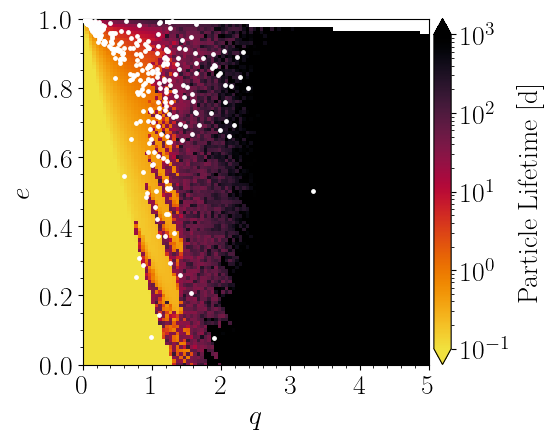}{0.25\textwidth}{Case 15}
        \fig{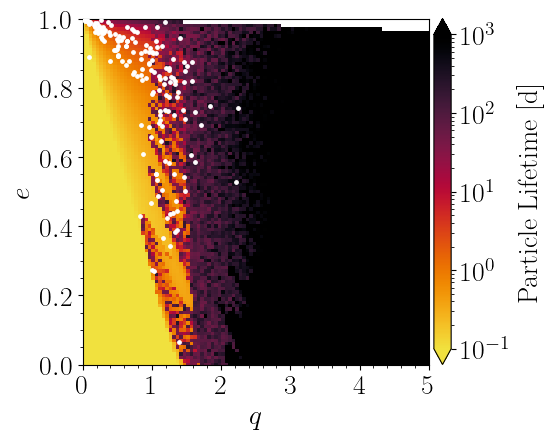}{0.25\textwidth}{Case 16}
    }
    \caption{\label{fig:label} Temporary satellites plotted in gray on a colorscale of dynamical lifetimes as a function of orbital eccentricity and pericenter distance. The impact properties for each case can be found in Table \ref{table:1}.}

\clearpage

\end{figure}
\begin{figure}[hp]
    \gridline{
        \fig{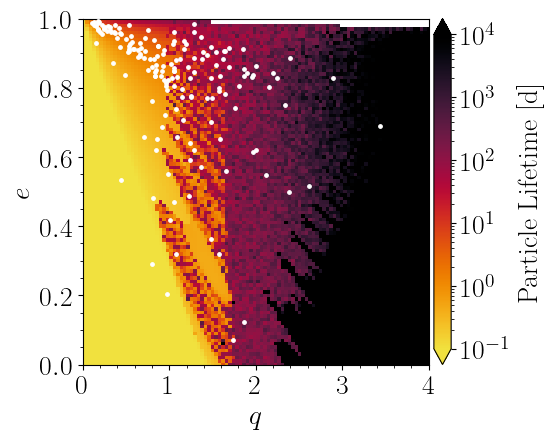}{0.25\textwidth}{Case 17} 
        \fig{stability_plots/lifetime_5_60_13_NoSpin_18deg.png}{0.25\textwidth}{Case 18}
        \fig{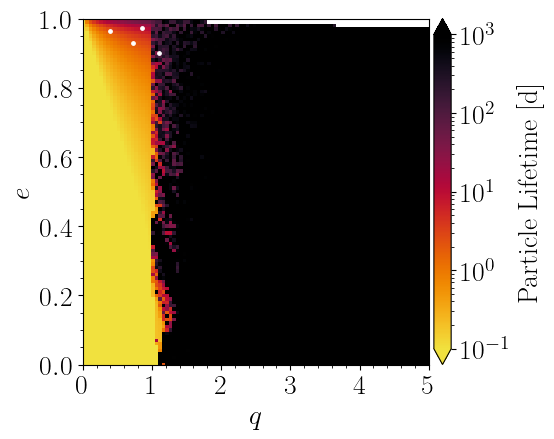}{0.25\textwidth}{Case 19}
        \fig{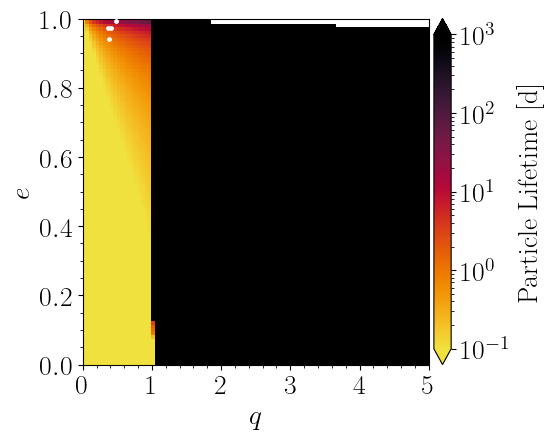}{0.25\textwidth}{Case 20}
    }

    \gridline{
        \fig{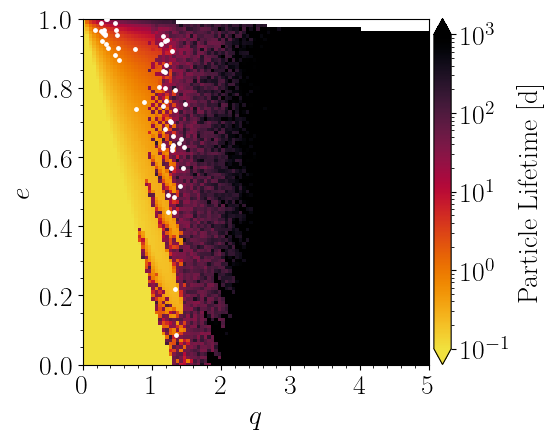}{0.25\textwidth}{Case 21} 
        \fig{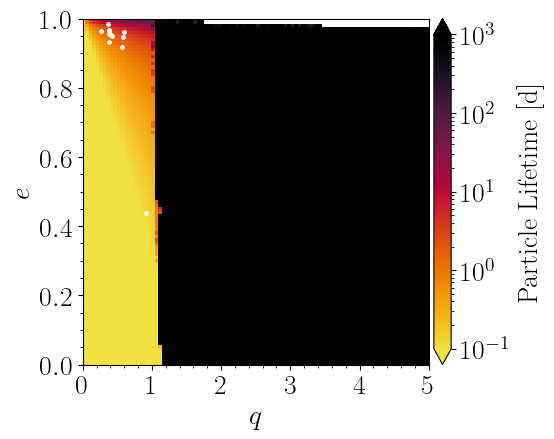}{0.25\textwidth}{Case 22}
        \fig{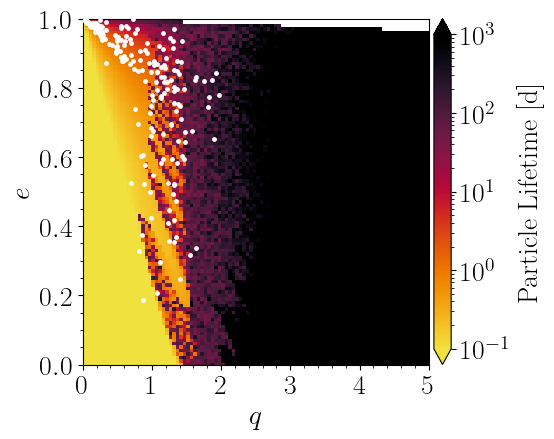}{0.25\textwidth}{Case 23}
        \fig{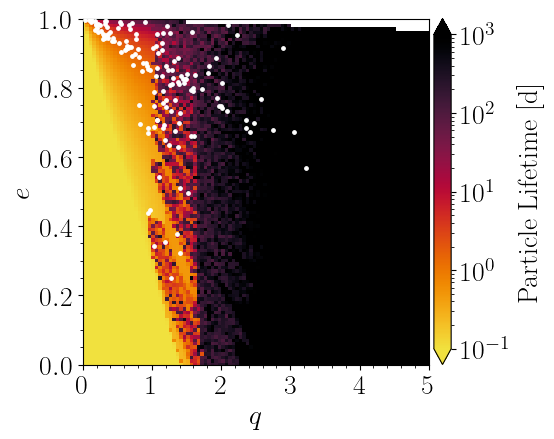}{0.25\textwidth}{Case 24}
    }

    \gridline{
        \fig{stability_plots/lifetime_5_60_13_NoSpin.png}{0.25\textwidth}{Case 25} 
        \fig{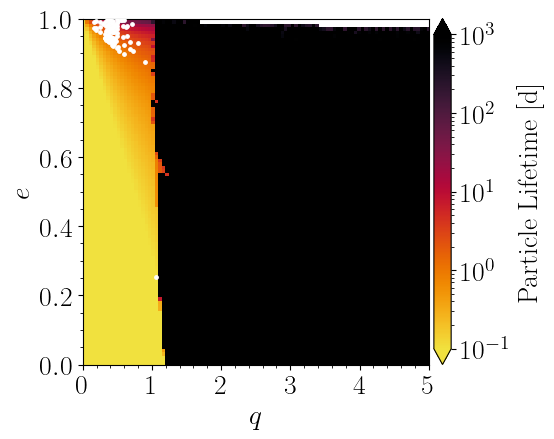}{0.25\textwidth}{Case 26}
        \fig{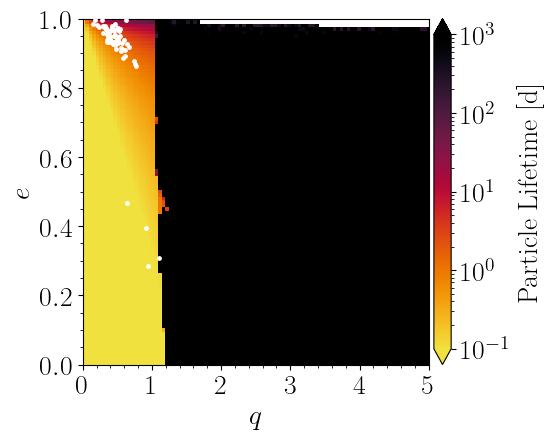}{0.25\textwidth}{Case 27}
        \fig{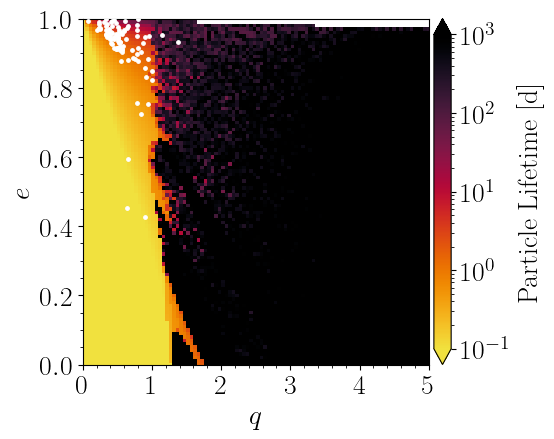}{0.25\textwidth}{Case 28}
    }

    \gridline{
        \fig{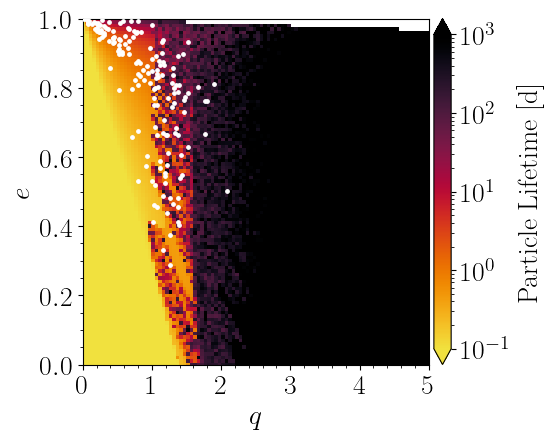}{0.25\textwidth}{Case 29} 
        \fig{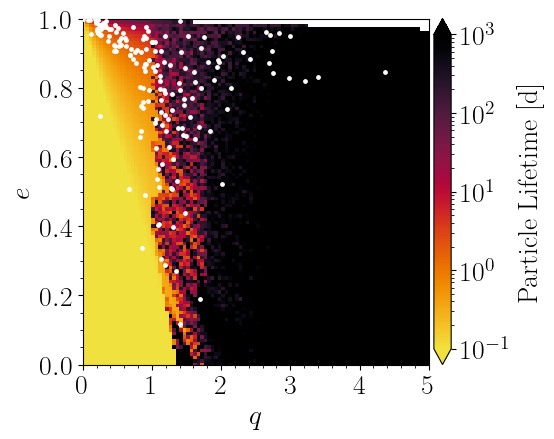}{0.25\textwidth}{Case 30} 
        \fig{stability_plots/lifetime_5_60_13_NoSpin_NoFriction.png}{0.25\textwidth}{Case 31}
        \fig{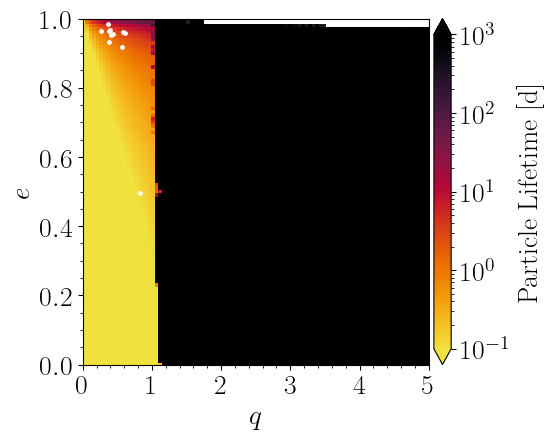}{0.25\textwidth}{Case 32}
    }
    \caption{Temporary satellites plotted in gray on a colorscale of dynamical lifetimes as a function of orbital eccentricity and pericenter distance. The impact properties for each case can be found in Table \ref{table:1}.}
\end{figure}

\clearpage

\bibliography{Walsh_bigbinary}{}
\bibliographystyle{aasjournal}



\end{document}